\documentclass{aa}  

\usepackage{natbib}
\bibpunct{(}{)}{;}{a}{}{,} 

\usepackage{graphicx}
\usepackage{txfonts}
\usepackage{multirow}
\usepackage{longtable}
\usepackage{pdflscape}

\usepackage{xspace} 
\renewcommand{\AA}{\normalfont\r{A}\xspace} 
\newcommand{\teff}{\ensuremath{T_{\mathrm{eff}}}\xspace}
\newcommand{\kms}{\ensuremath{\rm{km}\,s^{-1}}\xspace}
\newcommand{\logg}{\ensuremath{\log g}\xspace}
\newcommand{\feh}{\ensuremath{\rm{[Fe/H]}}\xspace}

\usepackage{xcolor,ulem}
\newcommand\hl{\bgroup\markoverwith{\textcolor{yellow}{\rule[-.5ex]{2pt}{2.5ex}}}\ULon} 

\begin{document}

   \title{Stellar atmospheric parameters for 754 spectra from the X-shooter Spectral Library\thanks{Full Tables A.1 -- A.3 are only available in electronic form at the CDS via anonymous ftp to cdsarc.u-strasbg.fr (130.79.128.5) or via http://cdsweb.u-strasbg.fr/cgi-bin/qcat?J/A+A/627/A138 }$^,$\thanks{Based on ESO observations from run IDs 084.B-0869(A/B), 085.B-0751(A/B) and 189.B-0925(A/B/C/D).}}

   \author{Anke Arentsen \inst{1,2}
          \and
          Philippe Prugniel \inst{3}
          \and
          Anais Gonneau \inst{4,2,5}
          \and
          Ariane Lan\c con \inst{5}
          \and
          Scott Trager \inst{2}
          \and
          Reynier Peletier \inst{2}
          \and
          Mariya Lyubenova \inst{6,2}
          \and
          Yan-Ping Chen \inst{7}
          \and
          Jes\'us Falc\'on Barroso \inst{8,9}          
          \and
          Patricia S\'anchez Bl\'azquez \inst{10}
          \and
          Alejandro Vazdekis \inst{8,9}
     }

   \institute{Leibniz-Institut f\"ur Astrophysik Potsdam (AIP), An der Sternwarte 16, D-14482 Potsdam, Germany \\
              \email{aarentsen@aip.de}
          \and
                Kapteyn Astronomical Institute, University of Groningen, Postbus 800, 9700 AV, Groningen, the Netherlands
         \and
              Centre de Recherche Astrophysique de Lyon (CRAL, CRNR, UMR 5574), Universit\'e Lyon 1, Ecole Nationale Sup\'erieure de Lyon, Universit\'e de Lyon
         \and
               Institute of Astronomy, University of Cambridge, Madingley Road, Cambridge CB3 0HA, United Kingdom
         \and
                Observatoire Astronomique, 11, rue de l'Universit\'e, F-67000 Strasbourg, France
        \and
                European Southern Observatory, Karl-Schwarzschild-Strasse 2, D-85748, Garching, Germany
        \and
                New York University Abu Dhabi, P.O. Box 129188, Abu Dhabi, United Arab Emirates
        \and 
                Instituto de Astrof\'isica de Canarias, V\'ia L\'actea s/n, E-38205 La Laguna, Tenerife, Spain
        \and
                Departamento de Astrof\'isica, Universidad de La Laguna, E-38205 La Laguna, Tenerife, Spain
        \and
                Departamento de F\'isica Te\'orica, Universidad Aut\'onoma de Madrid, 28049 Cantoblanco, Spain
             }

   \date{Received 18 September 2018 / Accepted May 3, 2019}

  \abstract
   {The X-shooter Spectral Library (XSL) is an empirical stellar library at medium spectral resolution covering the wavelength range from $3000$ \AA~to $24\,800$ \AA. This library aims to provide a benchmark for stellar population studies.}
   {In this work, we present a uniform set of stellar atmospheric parameters, effective temperatures, surface gravities, and iron abundances for  754 spectra of 616 XSL stars. }
   {We used the full-spectrum fitting package ULySS with the empirical MILES library as reference to fit the ultraviolet-blue (UVB) and visible (VIS) spectra. We tested the internal consistency and we compared our results with compilations from the literature.}
   {The stars cover a range of effective temperature $2900$ $< \teff < 38\,000$~K, surface gravity $0 < \logg < 5.7$, and iron abundance $-2.5 < \feh < +1.0$, with a couple of stars extending down to \feh $= -3.9$. The precisions of the measurements for the G- and K-type stars are 0.9\%, 0.14, and 0.06 in \teff, \logg, and \feh, respectively. For the cool giants with $\logg < 1$, the precisions are 2.1\%, 0.21, and 0.22, and for the other cool stars these values are 1\%, 0.14, and 0.10. For the hotter stars ($\teff > 6500$~K), these values are 2.6\%, 0.20, and 0.10 for the three parameters.
   } 
   {}


   \maketitle

   
\section{Introduction}

Libraries of stellar spectra are important resources in astronomy.
They are used in stellar population models for the analysis of galaxies and star clusters, and they are used as templates of stellar spectra for example for the automatic determination of stellar atmospheric parameters or for the validation of synthetic stellar spectra. For each of these purposes it is important to have spectra of as many different types of stars as possible and, especially for stellar population modelling, a large spectral range is preferable. 

The X-shooter Spectral Library (hereafter XSL; \citealt{chena}) is a stellar spectral library consisting of a collection of medium resolution spectra for 668 stars spanning a wide range of effective temperature (\teff), surface gravity (\logg, where g is expressed in cm s$^{-2}$) and iron abundance (\feh $ = \log(N_\mathrm{Fe}/N_\mathrm{H})_* - \log(N_\mathrm{Fe}/N_\mathrm{H})_{\odot}$ where N is the number density). The latter two parameters are written without units throughout.
The spectra were observed with the X-shooter instrument \citep{vernet} mounted at the Very Large Telescope (VLT), at a resolving power $R = \lambda / \Delta \lambda$ of $\sim 10~000$ in the wavelength range $\lambda = 3000$-$24~800$ \AA, where $\Delta \lambda$ is the full width at half maximum of the line-spread function (LSF).

With XSL, it will be possible to build the first stellar population models fully consistent from the ultraviolet to the near infrared.
Before XSL, in order to cover this wavelength region, the models had to combine different libraries made of different stars, observed at different spectral resolution, as for example in \citet{vazdekis16} or in \citet{villaume}. The resulting potential discontinuities limit the reliability and consistency of the predictions of the global spectral energy distributions.
For other applications, the large wavelength coverage will also be valuable to validate synthetic stellar spectra, or to cross-calibrate surveys made in different, narrower spectral regions.

Knowledge of the atmospheric parameters is required to use a library either for stellar characterization or for population models. In the latter case, for instance, the models combine individual spectra according to the expected number of stars in each cell of the parameter space (\teff, \logg, \feh).
The accuracy of the atmospheric parameters is in fact a recurring issue. For example, a modest bias of the temperature of the giant branch can change the determination of the age of old populations by several gigayear \citep[e.g.][]{charlot,koleva07} and offsets in the three parameters can mimic abundance ratio variations \citep{percival}.

Different methods can be used to determine stellar atmospheric parameters. At medium resolution, a well-established approach is full-spectrum fitting using synthetic or empirical libraries. The full-spectrum fitting method consists of comparing an observed spectrum to a template spectrum from a reference library that can be either empirical or synthetic. After adjusting the continuum and line broadening, the minimization of the residuals enables the estimation of the atmospheric parameters. The greatest advantage of full-spectrum fitting is that it allows the use of the whole spectral range and many spectral features at the same time, instead of picking specific features, therefore making optimal usage of the available signal. Some examples of full-spectrum fitting codes are TGMET (for Temperature, Gravity, MEtallicity, \citealt{katz}, \citealt{soubiran}), MATISSE \citep[MATrix Inversion for Spectral SynthEsis,][]{recio} and ULySS \citep[University of Lyon Spectroscopic Software,][]{koleva}.

Synthetic libraries, computed ab initio using stellar atmosphere models and synthesis of the spectral lines, provide references at any desired resolution and are free of noise or interference with peculiarities of real individual stars. Comparing synthesized and observed spectral lines is the only method to estimate the detailed abundances of stars. Therefore, synthetic libraries are a cornerstone of the characterization of stars. However, synthetic spectra cannot presently reproduce all the observed features of a stellar spectrum consistently across wavelength and parameter space \citep[e.g.][]{martinscoelho,coelho14}. This is because of uncertainties and approximations in the treatment of the physics and because of an imperfect knowledge of the millions of atomic and molecular transitions needed to predict the spectral lines \citep[see e.g.][]{kurucz17}. Some lines are not predicted at accurate enough wavelength and strength, other lines may simply be missing in the models.
On the other hand, empirical libraries can reproduce an observed spectrum with a precision of a few tenths of a percent \citep{wu}, although these libraries suffer from a number of observational defects, are affected by the peculiarities on their stars (such as rotation, binarity, or particular abundance patterns), and their parameter space coverage is restricted. This precision is a valuable advantage over using synthetic spectra at medium spectral resolution and in the regime of low temperatures, such as for late-K and M type stars, where isolated and well-modelled lines cannot be cherry-picked.

The goal of this paper is to provide homogeneous atmospheric parameters of as many XSL stars as possible. For XSL, using full-spectrum fitting with empirical references appears to be the most appropriate choice because of its spectral resolution and because of its large coverage of spectral type, in particular for cool stars. In this paper we analyse the XSL spectra with the ULySS package\footnote{\url{http://ulyss.univ-lyon1.fr}} \citep{koleva}, using the MILES empirical library \citep{sanchez,falconbarroso} as a reference. The spectrum to be analysed is compared to that computed with the MILES spectral interpolator \citep{prugniel11,sharma}, and the minimization of the residuals leads to an estimation of the atmospheric parameters. The MILES library (resolving power $R \sim 2200$) was designed to cover the needs of population synthesis applications and has a wide coverage of the parameter space. Its atmospheric parameters are bound to literature compilations \citep[e.g. PASTEL;][]{soubiran16} and homogenized as in \citet{cenarro}.

In Sec.~\ref{sec:data} we describe the current XSL spectral library and in Sec.~\ref{sec:method} we describe the method we use to determine the parameters. In Sec.~\ref{sec:param} we present the results and quantify the uncertainties, and we end with a summary in Sec.~\ref{sec:summary}.

\section{Data}\label{sec:data}

The spectra for XSL were obtained under an ESO Large Programme (run IDs 084.B-0869(A/B), 085.B-0751(A/B) and 189.B-0925(A/B/C/D)) at the VLT with the X-shooter instrument \citep{vernet}. The instrument splits the light using dichroics towards three echelle spectrographs: the ultraviolet-blue arm (UVB) covers the wavelength range $3000 - 5500$ \AA, the visible (VIS) covers $5300 - 10200$ \AA, and the infrared (NIR) covers $10000 - 24800$ \AA. The mean resolving powers of the spectral arms are $\sim 9000$, $11000,$ and $8000$ for the UVB, VIS, and NIR arms, respectively. 

The present work is based on a preliminary version of the second XSL data release (DR2: Gonneau et al., in prep). The DR2 consists of $\sim$800 UVB, VIS, and NIR spectra of 668 stars, and it includes the re-reduced spectra of DR1. Differences between the version used in this work and that to be released in the final DR2 concern only details of the flux calibration and have no significant effect on the measured parameters. The DR2 does not contain merged spectra of the three arms. In this paper, we analyse separately the UVB and VIS arms.

The spectra have associated error spectra obtained by propagating the estimated noise (photon noise plus detector noise) through the data reduction process. Some spectra of cool stars have very low signal in the blue. The minimum wavelength that can reasonably be considered as usable (keyword WMIN) is indicated in the DR2 and we used this information to trim the blue end of the spectra in our analysis.

\begin{figure}
\centering
\includegraphics[width=\hsize]{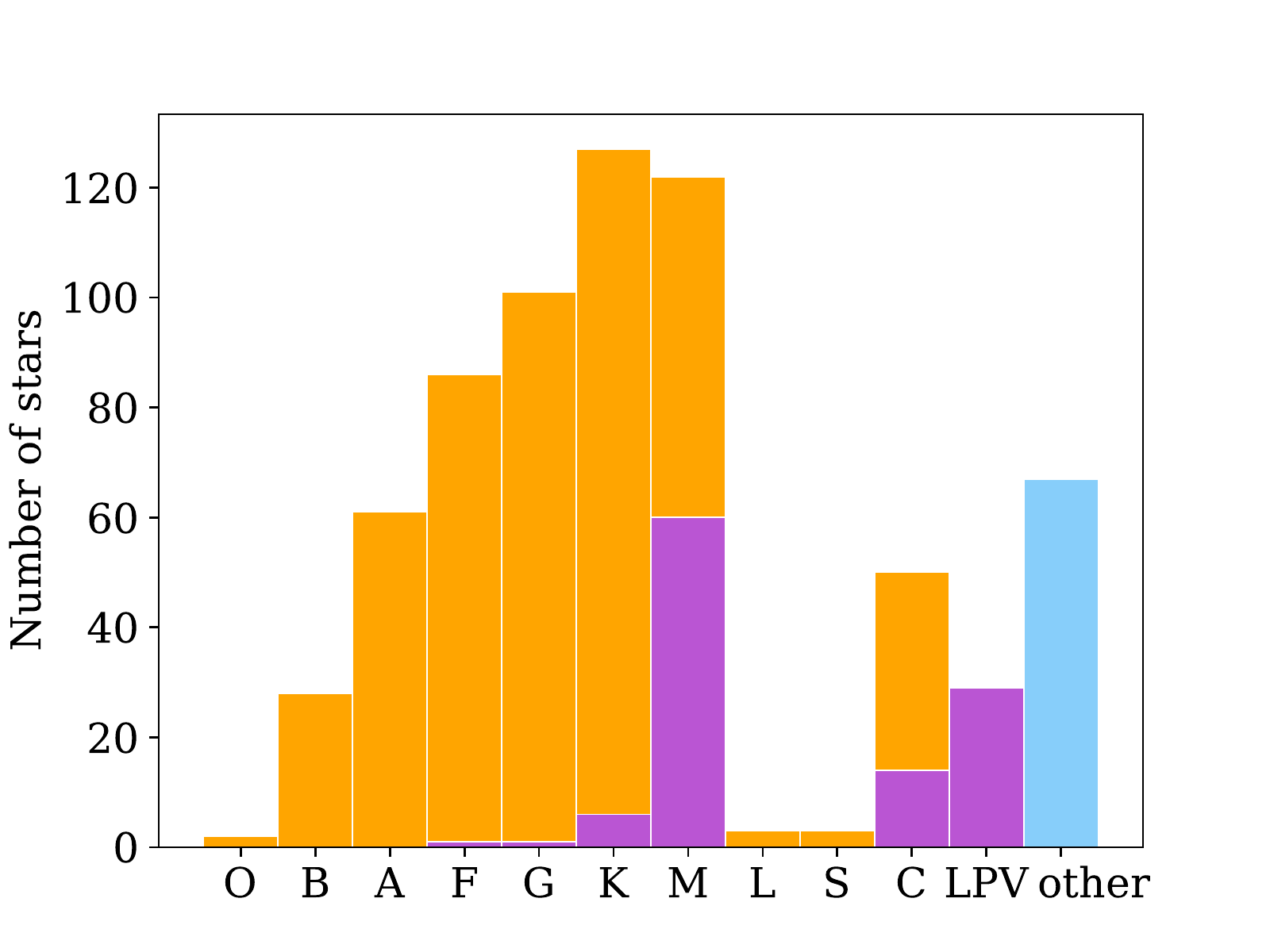}
\caption{Distribution of the spectral types present in XSL, as retrieved from SIMBAD. We highlight the number of stars known as LPV objects in purple. The stars in the `LPV' bin have no given spectral type but have object types labelling them as LPV objects. The stars in the blue bin `other' do not have spectral types in SIMBAD, nor have object types which label them as LPV objects.
          }
    \label{fig:hist}
\end{figure}

The details of the selection of stars for XSL will be described in the DR2 paper. In short, the XSL stars were selected to cover as much of the Hertzsprung-Russell diagram as possible; these stars have a wide range of metallicities. We did not attempt to provide our own spectral types based on the spectra. We retrieved the spectral types for as many stars as possible from the SIMBAD database \citep{simbad}, and we show the distribution in Fig.~\ref{fig:hist}. We highlight the long period variable (LPV) stars in each of the bins in purple. Some stars do not have a spectral type in SIMBAD, but have an object type which labels the star as LPV. We included these stars in the LPV bin of the figure. Additionally, there are stars that fall into the `other' category; these stars either do not have a spectral type and object type or they have one that labels them as a peculiar star (other than LPV). 

Some stars are observed more than once. Several cool giant stars were re-observed to monitor their spectral variability, some stars were re-observed to assess the stability of the instrument and data reduction, and others were repeated because of (sometimes unfounded) concerns about the observing conditions.

\section{Method}\label{sec:method}

We used ULySS (\citealt{koleva}, version 1.3.1) to determine the stellar atmospheric parameters for the spectra in XSL. This full-spectrum fitting package performs a $\chi^2$ minimization between a model spectrum and an observed spectrum. The model spectrum $S(\lambda)$ in ULySS is described by
\begin{equation} \label{eqn:ulyss}
    S(\lambda) = P_n(\lambda) \times G(v_{\mathrm{r}}, \sigma) \otimes \mathrm{TGM}(\teff, \logg, \feh,\lambda),
\end{equation} 

\noindent where $P_n(\lambda)$ is a series of Legendre polynomials up to degree $n$, and $G(v_{\mathrm{r}}, \sigma)$ is a Gaussian broadening function described by the radial velocity $v_{\mathrm{r}}$ and the broadening width\footnote{When analysing a line-of-sight integrated galaxy spectrum, $\sigma$ is the so-called velocity dispersion, while in the present case it is related to the rotational velocity. To some extent, G also absorbs uncertainties in the wavelength calibration and width of the LSF.} $\sigma$. The TGM function returns a model spectrum for given stellar atmospheric parameters \teff, \logg, and \feh by interpolating a reference stellar spectral library. The spectra are binned in logarithmic wavelength, so that the kinematical effects (Doppler shift and rotational broadening) can be expressed by a single convolution to all pixels. The  function $P_n(\lambda)$ corrects for the Galactic extinction and for any suboptimal flux calibration of the observed spectrum or the reference library.
The free parameters in the fit of an observed spectrum to a model spectrum are \teff, \logg, \feh, $v_{\mathrm{r}}$, $\sigma$, and the coefficients of $P_n(\lambda)$, which are all fit at the same time to deal consistently with degeneracies between the parameters.

The spectral interpolator approximates each wavelength bin of a spectrum with a polynomial function of \teff, \logg, and \feh. Each term of the polynomial is a product of powers of the three parameters \citep[see e.g. Sec. 2.5.2 in][]{prugniel11}. 
The interpolator combines three sets of polynomials corresponding to three temperature regimes, for OBA, FGK, and M stars, defined as follows: \[
      \begin{array}{lp{0.8\linewidth}}
            \mathrm{OBA~regime}: \teff > 7000 \mathrm{\,K} \\
            \mathrm{FGK~regime}: 4000 < \teff < 9000 \mathrm{\,K}  \\
            \mathrm{M~regime}: \teff < 4550 \mathrm{\,K} 

      \end{array}
      \]
      The three regimes are smoothly connected. For the hot stars the polynomials have 19 terms, and they have 26 terms for the warm and cool stars.

      Several spectral interpolators have been constructed by members of our group over time for the different popular stellar libraries, and some improvements have emerged with the successive versions. Because the results obtained with the MILES interpolator used in this work are not independent of those obtained with previous versions, we provide a brief historical summary. The first interpolator was made at a resolving power $R = 10\,000$ for the ELODIE library \citep{prugnielandsoubiran}. The library and the interpolator were later improved, and the current version 3.2 was described in \citet{wu}. The latter interpolator is available on the ULySS website. The ELODIE interpolator covers the wavelength range $3900 - 6800$~\AA and uses input atmospheric parameters compiled from the literature; each individual published set of measurements was carefully examined to adopt a ``best'' set for each star. The self-inversion of the library, i.e. the redetermination of the stellar parameters by fitting the library spectra with its associated interpolator allows for a check of the self-consistency of the input parameters and in fact is used to detect and correct inaccurate measurements in the compilation \citep[see e.g.][]{prugniel11}. The self-inverted parameters have a high internal consistency, which is a characteristic lacking in the heterogeneous literature compilation, but these self-inverted values are affected by the systematics of the interpolator. For example, the inverted parameters in some specific regions of the parameter space may display biases because the polynomials do not perfectly reproduce the physical variation. The choice of the terms in the polynomials, the definition of the temperature regimes, and the weighting of individual spectra were tuned to minimize those systematics.

      We note that the polynomials do not strictly interpolate between the spectra of the library, but also perform some approximation. This is an important aspect for empirical libraries as it results in a smoothing of both the noise and the effects of ``hidden'' parameters. These hidden parameters include all the characteristics of the stars that are not reflected by the three fundamental parameters. They are for example the individual abundances, the rotation, chromospheric activity, binarity (although binaries are as far as possible avoided in spectral libraries), and also include flux calibration errors, or uncertainty in the correction of the Galactic extinction.

      The polynomials naturally go wild outside the populated region of the parameter space. To alleviate the consequences, the interpolators therefore include extrapolation support, which is based on synthetic spectra differentially corrected to smoothly connect with the library (see \citealt{wu}).       
      
      Whereas the ELODIE interpolator was based on a literature compilation corrected for inconsistencies using a self-inversion, the MILES interpolator started from a catalogue of parameters obtained by fitting the MILES spectra using the ELODIE interpolator \citep{prugniel11}. Additionally, the MILES version was supplemented by literature values in regions of the parameter space in which the ELODIE parameters are not reliable. This approach took advantage of the high internal consistency in the FGK regime, where the ELODIE library is well populated and the interpolator is very reliable, but avoided being affected by the ELODIE interpolator systematics in the outskirts of the parameter space. This input catalogue and the self-inversion were carefully checked against the literature compilation.
      
      The last version of the MILES interpolator \citep{sharma} improved the modelling of cool stars, in particular by supplementing MILES with more spectra of M dwarfs. This feature is important for XSL, which contains a number of cool stars that are out of the range of ELODIE. Therefore we chose to use the MILES interpolator for the present analysis and we used the ELODIE interpolator for comparison tests.
      Our choice of the MILES interpolator sacrifices the original spectral resolution of XSL, which is higher than MILES. However \citet{prugniel11} and \citet{koleva12} have shown that the resolution does not have a strong influence on the precision of the derived parameters. The MILES library also offers a wider wavelength range ($3500 - 7400$~\AA) than ELODIE, which is an advantage in this study. 

      The reliability of ULySS with these interpolators has been tested in various papers (in particular \citealt{wu,prugniel11,koleva12,sharma}) by comparing the results with datasets from the literature. The method is robust for FGK stars over a wide range of iron abundances ($-2.5 < \feh < 0.3$) and gravities ($0 < \logg < 5$). In this regime the ELODIE and MILES libraries are well populated and the associated stellar parameters are the most reliable.
      For hotter or cooler stars, and for extreme gravities or iron abundances, the situation is more complicated. There are both fewer stars in the libraries, their parameters are less securely determined in earlier detailed studies, and furthermore the physics of these stars often cannot appropriately be described by only the three parameters that we are using. For hot stars, in addition, the narrow and weak lines usually considered as markers of the chemical composition are typically smeared out at medium or low spectral resolution, hence affecting the possibility of measuring abundances; we note however that this effect has not been formally checked. Thanks to the careful tuning of the MILES interpolator, the method generally behaves well in an extended region of the parameter space, but comparisons with external measurements remain important to detect issues.

      The only types of stars included in XSL that cannot a priori be measured with our method are (i) late-M and L-type dwarfs, (ii) carbon stars, and (iii) other peculiar stars.
With the wavelength coverage from the MILES interpolator ($3500 - 7400$~\AA) we can analyse the UVB and VIS spectra in XSL. We choose to fit the UVB between $4000-5500$~\AA, except for the hottest stars for which we use the range $3800-5500$~\AA~to include a few more lines in these almost line-less spectra, and the VIS between $5600-7400$~\AA. Above 5500~\AA for the UVB and below 5600~\AA for the VIS the spectrum is too contaminated by the unstable response of the dichroic of the X-shooter instrument. 

\subsection{Resolution matching}

If we used a reference stellar library with a higher spectral resolution than XSL, the procedure would be to use the relative LSF\footnote{The relative LSF of spectrum A with respect to spectrum B is the broadening that shall be applied to spectrum B to match the resolution of spectrum A. In general, the LSF depends on the wavelength and it may depart from a Gaussian.} of XSL with respect to the library to degrade the resolution of the reference library to match that of XSL. In the present case we have to do the opposite.

\begin{figure}
\centering
\includegraphics[width=\hsize]{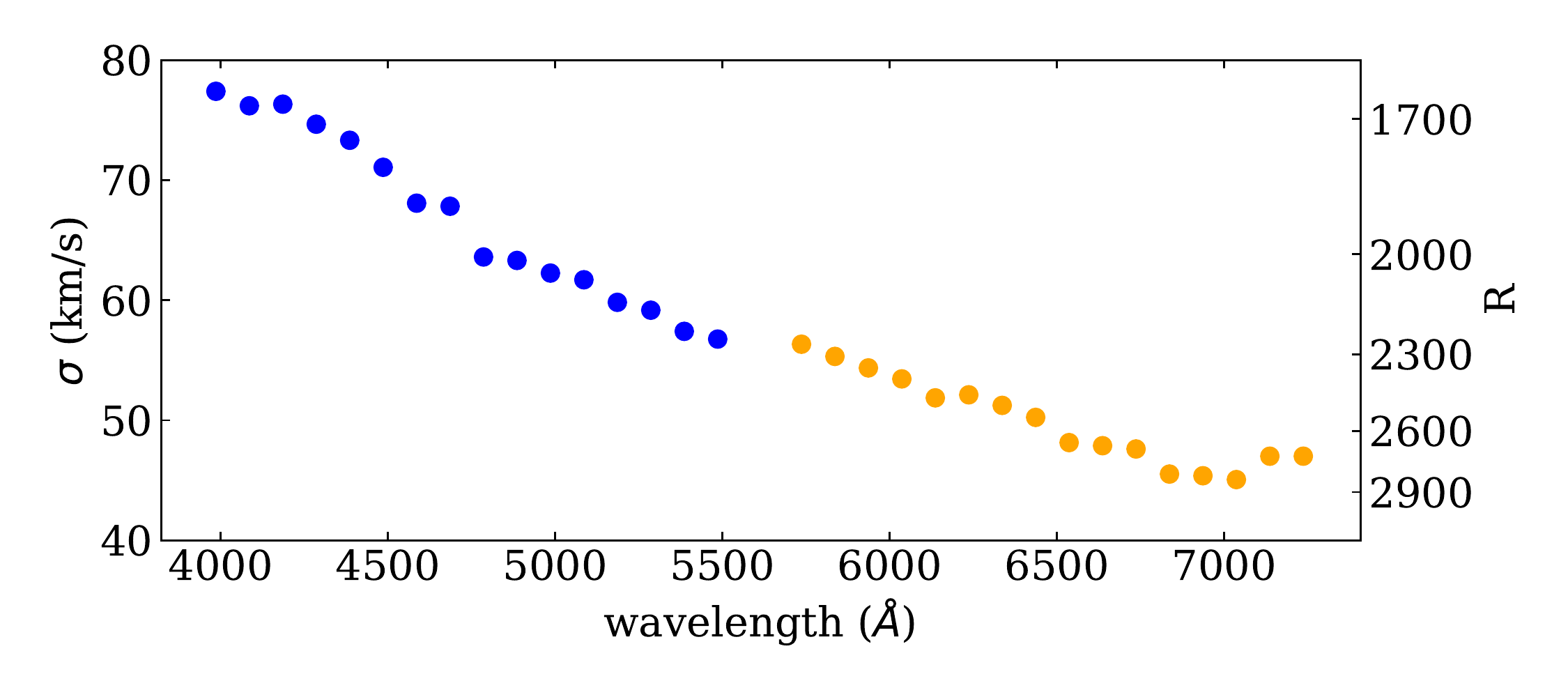}
\caption{Line spread function for the MILES interpolator, separated by colour for the UVB and VIS.}
    \label{fig:lsf}
\end{figure}

The LSF represents the broadening of spectral lines due to the instrumentation and data reduction. In the present paper we are assuming it is Gaussian, which is close to the actual shape, and the difference is not expected to influence the present analysis. The LSF of the MILES interpolator varies with wavelength, but as its determination in \citet{sharma} does not cover the full wavelength range used in this paper, we redetermined it by comparing an interpolated spectrum for the parameters of the Sun to the National Solar Observatory solar spectrum stored in HyperLeda\footnote{http://leda.univ-lyon1.fr/} \citep{makarov14} and to a high resolution synthetic spectrum. The resulting MILES interpolator LSF is represented in Fig.~\ref{fig:lsf}. 
Its Gaussian velocity dispersion varies from $\sim$80~\kms ($R \sim 1600$) at 4000~$\AA$ to $\sim$45~\kms ($R \sim 2800$) at 7000~$\AA$.

The XSL LSF, described in the DR2 paper, has a Gaussian velocity dispersion of about 13 \kms and 11 \kms, respectively, in the UVB and VIS regions that we are using. The variation and non-Gaussianity of the XSL LSF with wavelength can be neglected for our purposes. The velocity dispersion of the relative LSF of MILES with respect to XSL can be computed as $ \sigma^2_\mathrm{relative \,\,LSF} = \sigma^2_\mathrm{MILES \,\,LSF} - \sigma^2_\mathrm{XSL \,\,LSF}$. The relative LSF is different from the MILES LSF by only $\sim$2~\kms because $\sigma^2_\mathrm{XSL \,\,LSF}$ is much smaller than $\sigma^2_\mathrm{MILES \,\,LSF} $. However, a slight over-smoothing of the spectra is actually beneficial to the convergence of the radial velocities, therefore we simply convolve the XSL spectra with the MILES LSF instead of the relative LSF.

We used the ULySS function {\sc uly\_spect\_lsfconvol} to inject the MILES interpolator LSF in the XSL spectra. Because the resulting spectrum would be needlessly oversampled, we log-rebinned the spectra to a pixel value of
50 \kms , which is  comparable to the sampling of MILES (in the red spectral range, where MILES has the highest resolution, the pixel size is 2.25 times the full width at half maximum of the LSF). 

\subsection{Rest-frame reduction}

Although our full spectrum fitting method adjusts the velocity (parameter $v_{\mathrm{r}}$ in Eq.~\ref{eqn:ulyss}), this velocity shift should not exceed, by a couple of times, the broadening of the library. As XSL includes stars with radial velocities up to almost 400 \kms, it is necessary to reduce the spectra to the rest frame; this is achieved by changing the world-coordinate system of the spectra and does not require any additional rebinning. We corrected the spectra using the velocities which were determined as described in the DR2 paper.

\subsection{Multiplicative polynomial}\label{sec:mulpol}

The optimal value for the degree $n$ of $P_n(\lambda)$ mainly depends on the resolution, the fitted wavelength range, and the accuracy of the wavelength calibration of the spectra. We used the test described in \citet{koleva} to find the best value for \textit{n} for our spectra. We selected six stars of different spectral types and fit these with $n$ ranging from 0 to 300 in the wavelength region $4000 - 5500$~\AA for the UVB arm, and $5600-7400$~\AA for the VIS, from which we determined the value for $n$ where the parameters seem to become independent of $n$.

The solutions are nearly stable for $n \gtrsim 10$. When using echelle spectra, such as those from the ELODIE library or the present XSL spectra, the case for using higher values of $n$ comes from the presence of residuals from the blaze correction that may leave wavy fluctuations in the flux calibration with an amplitude of the order of one percent. For high S/N observations, this effect is prominently visible in the residuals of the fit with ULySS, although these residuals barely affect the solution because these patterns are not correlated with any physical feature. High values of $n$ correct this effect, but they sometimes tend to absorb physical features, in particular for hot stars for which the wings of broad H lines are partly fitted by the polynomial. This raises concerns that the exact degree of the polynomial may affect the solution. However, we analysed the spectra in this work both with $n=10$ and $n=30$, and it appears that taking a higher degree does not bias the parameters. Stars between $\sim$7000--9000~K are an exception; for these stars the resulting \teff from the UVB fit clearly depends on the choice of $n$. We estimate the bias introduced in Sec.~\ref{sec:uvbvis}.

With $n=30$, most of the echelle residuals and features of imperfect flux calibrations are taken out. 
Therefore we adopt $n=30$ as the degree for $P_n(\lambda)$, for both the UVB and the VIS spectra. For some low S/N spectra for which the blue sections of our wavelength range are below the S/N threshold, $n$ is reduced proportionally to the wavelength range actually used.

If the multiplicative polynomial goes to negative values or ULySS cannot find a fit, ULySS automatically reduces $n$. This mainly happens for some of our cool giant stars.

\subsection{Fitting a spectrum}\label{sec:fitspec}

The ULySS package performs a local minimization, starting with a set of parameter guesses. The convergence region is generally very wide, meaning that starting with a guess very distant from the actual parameters, the program still converges to the right solution. However, in some cases the solution may be trapped in a local minimum. This mainly happens when the star lies in the margins of the parameter space. The solution to find the absolute minimum is to repeat the minimization starting from a variety of initial guesses, and adopt local solutions with the best $\chi^2$. In ULySS, this is easily achieved by providing a grid of guesses rather than a single value. We use the following grid: 
\[
      \begin{array}{lp{0.8\linewidth}}
            \teff \in [3000, 4000, 5600, 7000, 13000, 30000] \\
            \logg \in [0.5, 1.8, 3.8] \\
            \feh \in [-1.7, -0.3, 0.5]
      \end{array}
   \]

The bounds of the parameter space are set to [$2700, 40000$] K, [$-0.5, 5.9$], and [$-4.0, 1.0$] for \teff, \logg, and \feh, respectively. The complete algorithm is
\begin{itemize}
\item Mask emission lines that may prevent the convergence, or make it slower
\item Run the minimization with the above grid of guesses
\item Do another minimization starting from the best solution of the run with multiple guesses, using the \texttt{/CLEAN} option, which automatically rejects bad regions in the fit (owing to e.g. bad pixels, emission lines, and telluric lines) to improve the final solution.
\end{itemize}

We include an extra step for stars hotter than 8000~K, for which we found that including the \texttt{/CLEAN} option sometimes results in a bad fit indicated by unphysical $\sigma$ (broadening width) values. The reason for this is that hot stars only have a few spectral lines, and if they are cleaned out because they are not well fitted, a low $\chi^2$ is achieved by smoothing the spectra. Therefore, if $\sigma$ is higher than 500~\kms, we redo the fit without the \texttt{/CLEAN} option on. This results in a much better fit for many of these hot stars. 

\section{Results}\label{sec:param}

We fitted 814/815 UVB/VIS spectra of 668 unique stars included in XSL DR2 to obtain their stellar atmospheric parameters.
In this section we select the acceptable measurements, analyse the errors, and combine the measurements in the UVB and VIS arms.

\subsection{Selection of the reliable solutions}\label{sec:reliable}

In this section, we establish criteria to identify the cases in which the derived parameters are unreliable either because the observation has an insufficient quality (in the wavelength region used) or for other reasons. Our analysis method is valid only for stars that can be modelled with the interpolator. Some special stars in particular, such as carbon stars (but see \citealt{gonneau16, gonneau17} for a different approach) and other peculiar stars, cannot be modelled well. 

The reduced $\chi^2$ of the fit, which compares the magnitude of the residuals to the fit with the expected error from the noise spectrum, indicates the quality of the fit. But as the interpolator has a limited capability to represent real spectra, very high S/N spectra are unavoidably fitted with a large $\chi^2$ ; reciprocally, very low S/N observations may be fitted with  $\chi^2 \approx 1$, even if the parameters of the stars are not correctly matched. Therefore, the criterion cannot simply be to reject the solutions with $\chi^2$ larger than a threshold. A more realistic approach is to set a threshold on the signal to residuals ratio (SRR; computed over the whole wavelength range of the fit; normally SRR $<$ S/N). 

The fitted broadening width $\sigma$ (see Eq.~\ref{eqn:ulyss}) is related to the stellar rotation for some hot stars, but high values generally indicate a failed fit. When no proper match is found in the library, the minimum $\chi^2$ is obtained by strongly smoothing the model spectrum.

We empirically determined the following recipe to consider a solution as acceptable for the UVB arm:

\begin{itemize}
\item Reject solutions with $\sigma > 500$ \kms
\item Reject solutions with $\sigma > 300$ \kms and $\teff < 10\,000$ K
\item For $\teff < 3600$ K, reject solutions with SRR$ < 1.5$ 
\item For $\teff > 3600$ K, reject solutions with SRR$ < 15$
\end{itemize}

\noindent And for the VIS arm:
\begin{itemize}
\item Reject solutions with $\sigma > 300$ \kms
\item For $\teff < 3600$ K, reject solutions with SRR$ < 3$ 
\item For $\teff > 3600$ K, reject solutions with SRR$ < 25$
\end{itemize}

\noindent These criteria were tuned after visual inspection of the residuals to the fits to exclude the solutions that did not converge (they generally went to high $\sigma$) or where, because of the low quality of the spectra, we could not see any match of the spectral features. These criteria also rejected most of the carbon stars because the carbon features cause large residuals, and for those that were accepted, the non-carbon features are well fitted. Parameter estimations from 734 UVB and 737 VIS observations are considered reliable according to these criteria.

\subsection{Error analysis}\label{sec:erroranalysis}

The propagation of the errors from the noise in the observations provides us with error estimates on the parameters computed by the ULySS software. This is what we call the formal error (or precision). However, the error propagation through the complex data reduction and analysis may suffer from some approximations or inaccuracies, and therefore the formal errors may not be an unbiased estimator of the actual effect of the noise. Therefore, in Sect.~\ref{sec:intprec}, we use the repeated observations to correct the formal errors and derive what we call the internal errors.

It is well known that the internal errors in stellar parameter determination strongly underestimate the real errors. There are some clear reasons for this. Several observational effects (e.g. non-linearity of the detector and imperfect subtraction of diffuse light in the spectrograph), peculiarities of the individual stars, hidden parameters (e.g. abundance patterns, rotation, and binarity), and inaccuracies and systematics of the models (in this case the spectral interpolator) all contribute to the final error budget. These effects are at the origin of systematics between series of measurements obtained with different data or methods. They have been studied in a number of literature studies \citep[e.g.][]{Smiljanic2014}, and a working group of the International Astronomical Union has been tasked to propose practices to bring a better understanding of this error budget\footnote{\url{http://svo2.cab.inta-csic.es/projects/iaug5wg}}. In Sections~\ref{sec:systematics_interpolator} and \ref{sec:adopted} we complete our error analysis to estimate the total error.

Finally, our results are compared to some literature compilations in Section~\ref{sec:extcomp} and \ref{sec:previousulyss} to provide us with some hints of the external errors, although attributing the fair share of responsibility of the differences between our measurements and the used reference is mostly a subjective choice. 

\subsection{Internal precision}\label{sec:intprec}

The validity of the formal errors returned by ULySS relies on the estimated variance of each bin of the spectrum and on the correlation of the noise between successive bins\footnote{The rebinning of the spectrum results in a correlation of the noise over adjacent pixels which appear as non-diagonal terms in the co-variance matrix. The error spectrum is the square root of the diagonal of this matrix. In ULySS, this effect can be modelled using a factor that corrects the number of independent pixels; neglecting this correlation results in underestimating the errors.}. Using a long series of observations of the same star, \citet{sharma} showed that if the noise on the spectrum and its correlation are well estimated, the formal fitting errors on the derived parameters match the internal errors, i.e. the errors resulting solely from the noise.

In our sample we do not have one star with a long series of observations, but we do have many stars with multiple observations, observed either two, three, or four times. We use the difference in determined parameters between in total 135 pairs of observations that we consider reliable to estimate the validity of the formal errors returned by ULySS, where we have excluded stars with \teff < 3700 K, which are likely variable.

\begin{figure*}
\centering
\includegraphics[width=17cm]{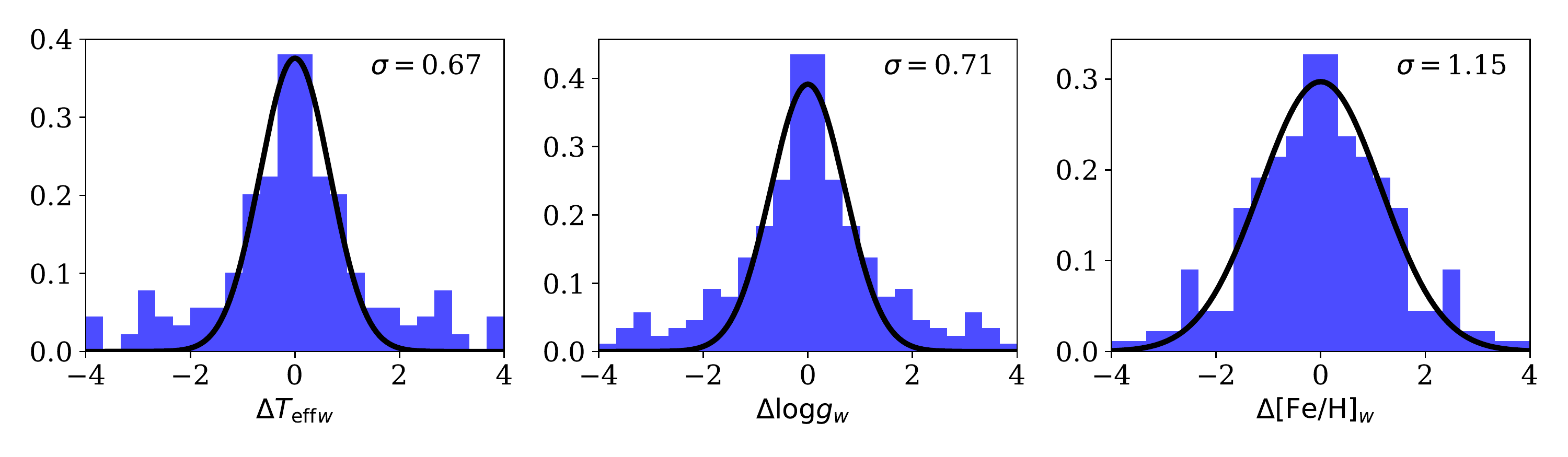}
\includegraphics[width=17cm]{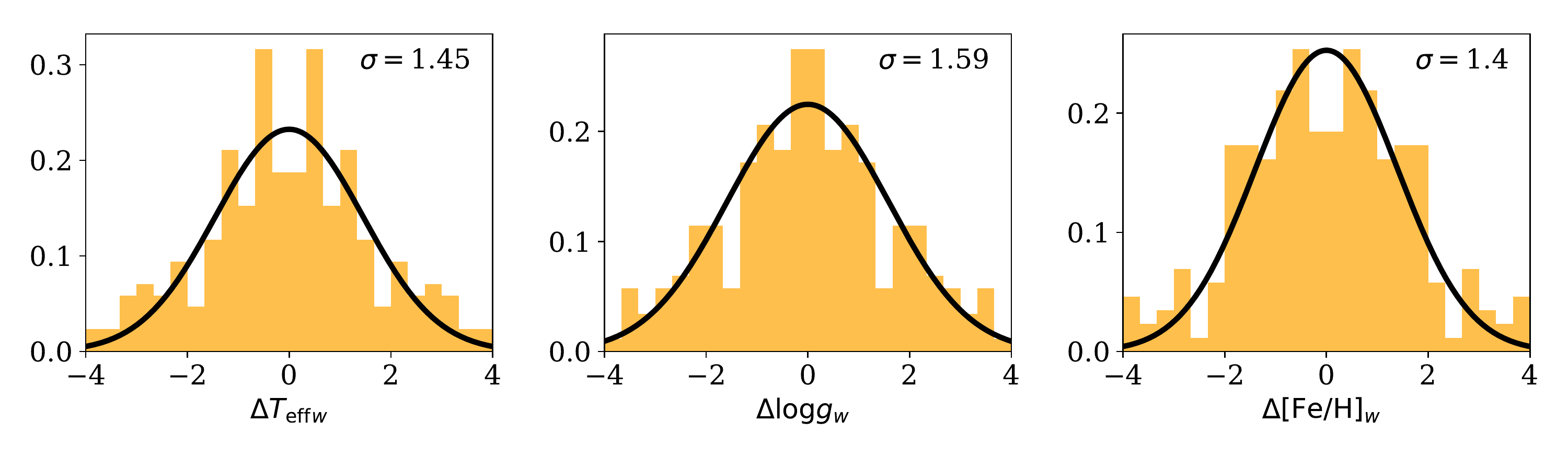}
\caption{Distributions of $\Delta \mathrm{P}_{w}$ for pairs of observations of the same star. The UVB and VIS analyses are presented in the top and bottom row, respectively. The Gaussians with a mean of zero and a standard deviation computed as described in the text are shown, the standard deviations of these Gaussians are given in the top right corner of each panel. }
    \label{fig:repeatedobs}
\end{figure*}

Not all spectra have the same S/N, therefore we compute the following weighted difference for each pair of observations $i$: 
\begin{equation}
    \Delta \mathrm{P}_{w,i} = \frac{\mathrm{P}_{1,i} - \mathrm{P}_{2,i}}{\sqrt{\epsilon_{1,i}^2 + \epsilon_{2,i}^2}},
    \label{eqn:P}
\end{equation} 

\noindent where P can be \teff, \logg or \feh, $\epsilon$ is the formal ULySS error on the respective parameters, and $1$ and $2$ indicate two observations of the pair. Statistically, it should follow a Student's t distribution, which is close to a Gaussian but has heavier tails. If the formal and internal errors are equal, we would expect a standard deviation of one for the distribution of $\Delta \mathrm{P}_w$. A deviation from one indicates that the estimated noise spectra may be incorrect. We computed standard deviations using the python \texttt{sigma\_clip} function with a clipping sigma of 2.2. 
We present the results in Fig.~\ref{fig:repeatedobs}, in which each pair is shown twice to produce symmetrical distributions.

The standard deviations are of the order of 0.7 for the UVB and of 1.5 for the VIS. Experimenting with the degree of the multiplicative polynomial (see Sect.~\ref{sec:mulpol}) we found that these standard deviations change by about 10\%. The prominent difference between the two arms is likely due to an incorrect estimate of the noise spectra in DR2. The suspected reasons are that the noise variance was interpolated rather than rebinned when the spectra were transformed into evenly distributed wavelength bins, and that we did not take into account the correlation of the noise. Both effects depend on the ratio of the size of the pixels on the detector and in the final spectrum, which varies with wavelength and between the arms. We did not investigate the effect further because it belongs to the data reduction. We can rule out that the arms contrast is from causes other than the noise propagation, for example to different sensitivity to degeneracy or systematics of the interpolator, because those effects would change significantly over the parameter space, which we do not observe. We therefore use the above factors to rescale the formal errors into internal errors.

We finally note that the variations in the stellar parameters between repeated observations of stars are very small. After $3\sigma$ clipping (which excludes 8 out of 135 pairs), the dispersion is 0.5\% for differences in \teff, 0.05 for differences in \logg, and 0.03 for differences in \feh.

\subsection{Systematics and total errors}\label{sec:systematics_interpolator}

In this section we investigate the systematic effects introduced by the spectral interpolators and the other effects that contribute to the total error.
First we compare the UVB solutions obtained with the MILES and ELODIE interpolators in the same wavelength range. The two interpolators were computed from two different libraries (different stars and spectra, and different distribution in the parameter space), and therefore we may expect different systematic effects. 

Then, we compare the analysis in the two arms with the MILES interpolator, which, as they correspond to two non-overlapping wavelength regions, can also be regarded as two different interpolators. Indeed, as each spectral bin is independently modelled with a specific polynomial, the interpolator in two spectral regions may have unrelated systematic effects, even if the library is the same. As two separate wavelength regions were used, the effect of hidden parameters, such as rotation, binarity, contamination by the light of nearby stars, or detailed abundances, are likely to be different.

While the first test addresses only the question of the systematics of the interpolators, the second is also sensitive to the other observational and physical effects. The second test is a more complete assessment of the total error.

\subsubsection{Comparison between MILES and ELODIE solutions (UVB)}

\begin{figure*}
\centering
\includegraphics[width=17cm]{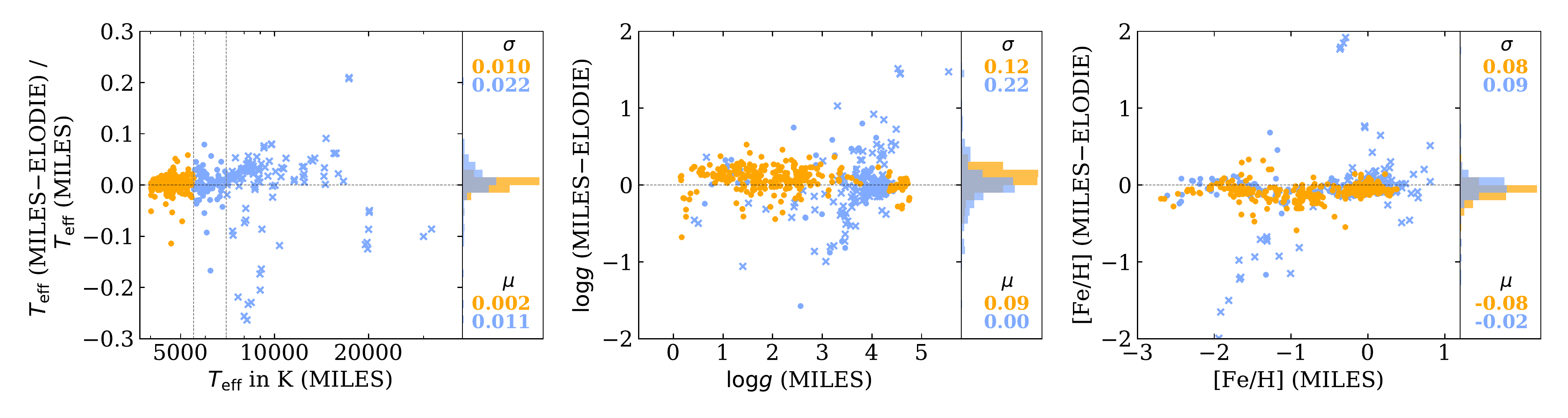}
\caption{Comparison between the MILES and ELODIE solutions in the UVB for the stars with \teff (MILES) $> 4000$ K. Each atmospheric parameter is shown in a separate panel, and the colour of the symbols identify two temperature ranges (dividing at 5500~K). Stars hotter than 7000~K are shown as crosses. In the \teff and \logg panels, seven and two outliers, respectively, are not shown (all hot stars). The dispersion $\sigma$ and mean offset $\mu$ for the two temperature ranges are indicated.}
    \label{fig:miles-elodie}
\end{figure*}

We computed the parameters again with the higher resolution ELODIE interpolator for our UVB spectral range; the ELODIE range is limited to 6800 \AA in the red, and therefore we cannot perform the test for our VIS range. We used the same procedure, except that in this case the resolution of XSL spectra did not have to be reduced and we only corrected for the changing of the XSL LSF with wavelength. The comparison with the parameters derived using the MILES interpolator is shown in Fig.~\ref{fig:miles-elodie}, i.e. for stars with \teff > 4000 K, since the ELODIE interpolator has not been optimized for cooler stars. The standard deviations for stars with $4000 < \teff < 5500$ K and $\logg > 1$ are 1\%, 0.12, and 0.08 for \teff, \logg, and \feh, respectively, after clipping 3$\sigma$ outliers. 

\begin{figure*}
\centering
\includegraphics[width=17cm]{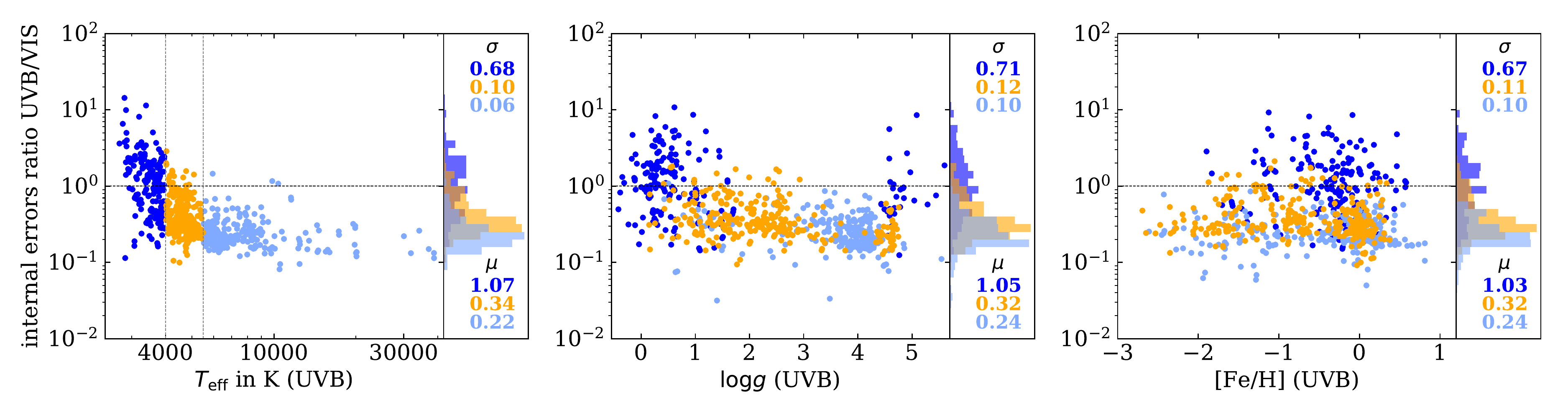}
\caption{Ratio of the internal errors between the UVB and VIS for each atmospheric parameter, in  separate panels. The stars with $\teff < 4000$ K, $4000 < \teff < 5500$ K, and $\teff > 5500$ K are shown with dark blue, orange, and light blue symbols, respectively. The dispersion $\sigma$ and mean offset $\mu$ for the three temperature ranges are indicated.}
    \label{fig:errors}
\end{figure*}

\begin{figure*}
\centering
\includegraphics[width=17cm]{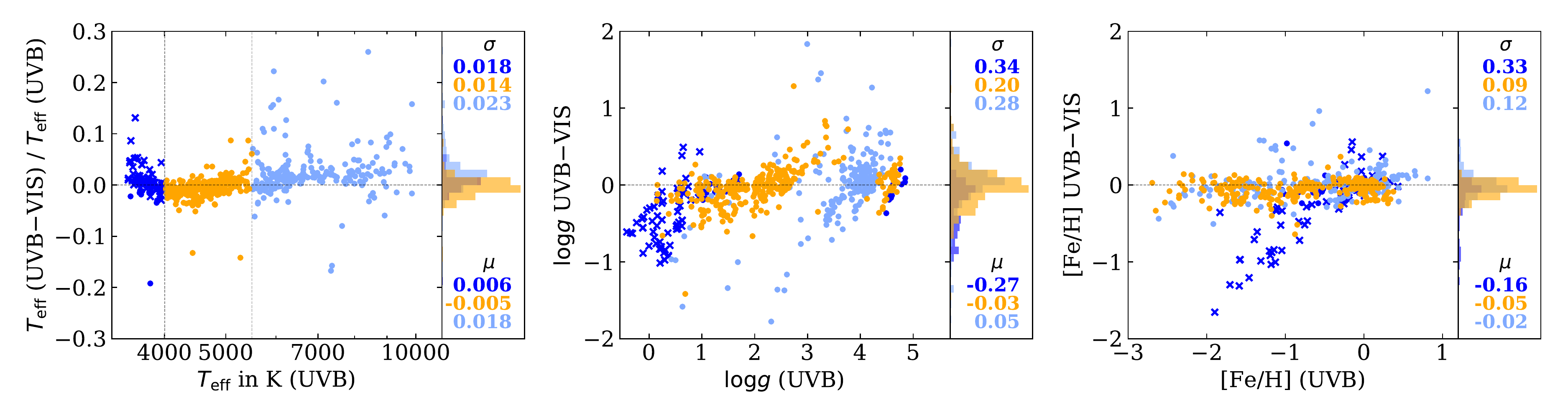}
\caption{Difference between the parameters determined from the UVB and  VIS spectra for stars with 3500 K $<$ \teff (UVB) $<$ 10\,000 K. The colour-coding is the same as in Fig.~\ref{fig:errors}, and for the coolest stars ($<$ 4000~K) the giants are shown with crosses. In the \logg panel, two outliers (hot stars) are not shown. The dispersion $\sigma$ and mean offset $\mu$ for the three temperature ranges are indicated.}
    \label{fig:UVBVIS}
\end{figure*}

There are several trends worth noting. For the coolest stars (\teff < 5500~K), the MILES \logg is systematically higher than that from ELODIE by 0.1, and the MILES \feh is systematically lower by 0.1. For $5500 < \teff < 7000$~K, the biases are smaller; MILES \logg is systematically higher by 0.05 and the MILES \feh is systematically lower than the ELODIE values by 0.05. Finally for the hottest stars (\teff > 7000~K) the most notable bias is in \teff, where the MILES values are on average 3\% higher. The magnitudes of these biases are comparable to the standard deviations of the difference between the two series. The MILES interpolator has been carefully tuned \citep{prugniel11,sharma} to minimize the biases between the parameters of the input catalogue and those obtained after a self-inversion of the library, and we may suppose that the biases observed are mostly due to the ELODIE interpolator. However, including the biases and the random dispersion, the $rms$ between the two series are 1.4\%, 0.18, and 0.11 for \teff, \logg, and \feh, respectively (after a 3$\sigma$ clipping), and if we conservatively assume that the systematics are equally shared between the two interpolators, the quoted dispersions shall be divided by $\sqrt{2}$ to estimate the magnitude of the errors introduced by the interpolators. Those are 1.0\%, 0.13, and 0.08 for the three parameters.

\subsubsection{Comparison between UVB and  VIS solutions}\label{sec:uvbvis}

In this section we investigate differences between the UVB and VIS solutions. We first compare the internal uncertainties from the fit to the UVB and VIS spectra.
In Fig.~\ref{fig:errors} we present for each observation the UVB internal error divided by the VIS internal error for the three atmospheric parameters.
For cool stars with $\teff < 5500$ K there is a decrease with temperature in the UVB/VIS internal errors ratio. This reflects the fact that the spectra of the coolest stars have lower S/N in the UVB compared to the VIS. For stars hotter than 5500 K the ratio of the UVB to VIS internal errors is constant and the UVB precisions are about four times higher. 
In Fig.~\ref{fig:UVBVIS} we compare the solutions from the UVB and VIS, differentiating the three \teff ranges $3500-4000$~K, $4000-5500$~K, and $5500-10000$~K with different colours. We are not comparing the solutions for the coolest and hottest stars because in the first case the VIS solution can a priori be considered superior because these cool stars have very little flux in the UVB, and in the latter case the UVB is likely superior because there are almost no lines in the VIS.
The two solutions appear essentially consistent, except these marginally significant biases: (i) the UVB \teff is higher than the VIS for the stars hotter than 5500 K, reaching a bias of 3\% at 10\,000 K; (ii) the VIS \teff is higher than the UBV by about 1\% for $3800 < \teff < 4800$ K; and (iii) the UVB \feh are systematically lower than the VIS \feh by about 0.045. A comparison with a literature compilation (see Sect.~\ref{sec:extcomp}) suggests that in the first case the bias is in the UVB solution, and in the two latter the bias is in the VIS solution. 
Our sample contains 161 spectra in the region of cool giants ($\teff < 4000$ K, and $\logg < 3$), represented as blue crosses in Fig.~\ref{fig:UVBVIS}. For the 129 of these with UVB solutions $\teff < 3800$ K, the UVB \feh are spread to low \feh (down to $-1.9$), while the VIS \feh is closer to solar. This produces the diagonal plume, with slope $-1$, stretching in the right panel, and for these spectra the UVB \logg are reaching extreme negative values. Many of these stars are cool Miras, and their flux in the UVB arm is low, so we tend to a priori dismiss those measurements. The comparison with the literature for 14 spectra of stars having high resolution estimates confirms that standard deviation between UVB and literature is higher than between VIS and literature. But it also appears that the VIS \feh are biased to higher values with respect to the literature by about 0.2.
We correct for these biases when we combine the two solutions in Sect.~\ref{sec:adopted}.
  
For 239 spectra with $4000 < \teff < 5500 $ K and $\logg > 1, $ the $rms$ deviations between the UVB and VIS solutions are 1.3\% (or 62 K), and 0.20 and 0.08 for \teff, \logg and \feh, respectively, after clipping the 3$\sigma$ outliers. Accordingly, estimated total errors are 0.9\% (or 44 K), 0.14 and 0.06 (the $rms$ divided by $\sqrt{2}$). This is comparable with the external errors quoted in \citet{wu} for the stars of spectral types F, G, and K: 43 K, 0.13 and 0.05 for the three parameters.

  For the cooler stars, $\teff < 4000$ K and $\logg > 1$ (58 spectra after clipping the outliers), the $rms$ deviations are 2.0\%, 0.19, and 0.13. The temperature and metallicity consistency degrades by about a factor 1.5. For the stars with $logg < 1$ (130 spectra after clipping the outliers), the deviations are 2.0 \%, 0.29, and 0.28. Finally, for the warmest stars, $6500 < \teff < 10\,000$ K (90 spectra after clipping the outliers), the deviations are 2.3\%, 0.30, and 0.12, i.e. 1.5 to 2 times less consistent than the GK stars. The consistency further degrades at higher temperature, down to 3.7\%, 0.29, and 0.13 for all 117 spectra warmer than 6500 K, including those warmer than 10\,000~K.
For the cool stars, the UVB measurements are less accurate than the VIS ones, and the opposite is true for the hot stars, in particular above 10\,000 K. Therefore, the comparison between the two arms may overestimate the total errors, but as we lack other indicators, we conservatively derive the total errors from these comparisons.

\subsection{Combination of the UVB and VIS solutions: Adopted parameters and total error}\label{sec:adopted}

The adopted parameters are finally derived by combining together the UVB and VIS solutions that satisfy the acceptance criteria stated in Sec.~\ref{sec:reliable}, after correcting the solutions for the biases described in Sec.~\ref{sec:uvbvis}. As we have seen above, and as we could logically expect, the UVB solution is generally better for the hot stars, and the VIS solution is better for the cool stars. We therefore adopted the following scheme:
\begin{itemize} 
\item Use the UVB solution alone if \teff (VIS) $> 4500$ K, or if the VIS solution is rejected.
\item Use the VIS solution alone if \teff (VIS) $< 10\,000$ K and either the UVB solution is rejected or if \teff (UVB) $< 3800$~K.
\item Otherwise use the average of the two solutions weighted by the inverse of the square of the internal errors.
\end{itemize}

We use the comparison between the UVB and VIS solutions to derive the total error. We computed statistics of the differences between the two solutions in different regions of the parameter space, assuming this represents the total errors. After subtracting quadratically the internal errors we derive models of the errors due to the interpolator and effects of the hidden parameters, and finally we combine these errors with the internal errors to estimate the total errors for each measurement.

We adopted one set of parameters for each observation, therefore, in the case a star has repeated observations, we have multiple sets of parameters for the star. As the stars may be variable, we do not attempt to combine these measurements. If necessary, averaging the different values may be the best approach.

\begin{figure}
\centering
\includegraphics[width=\hsize]{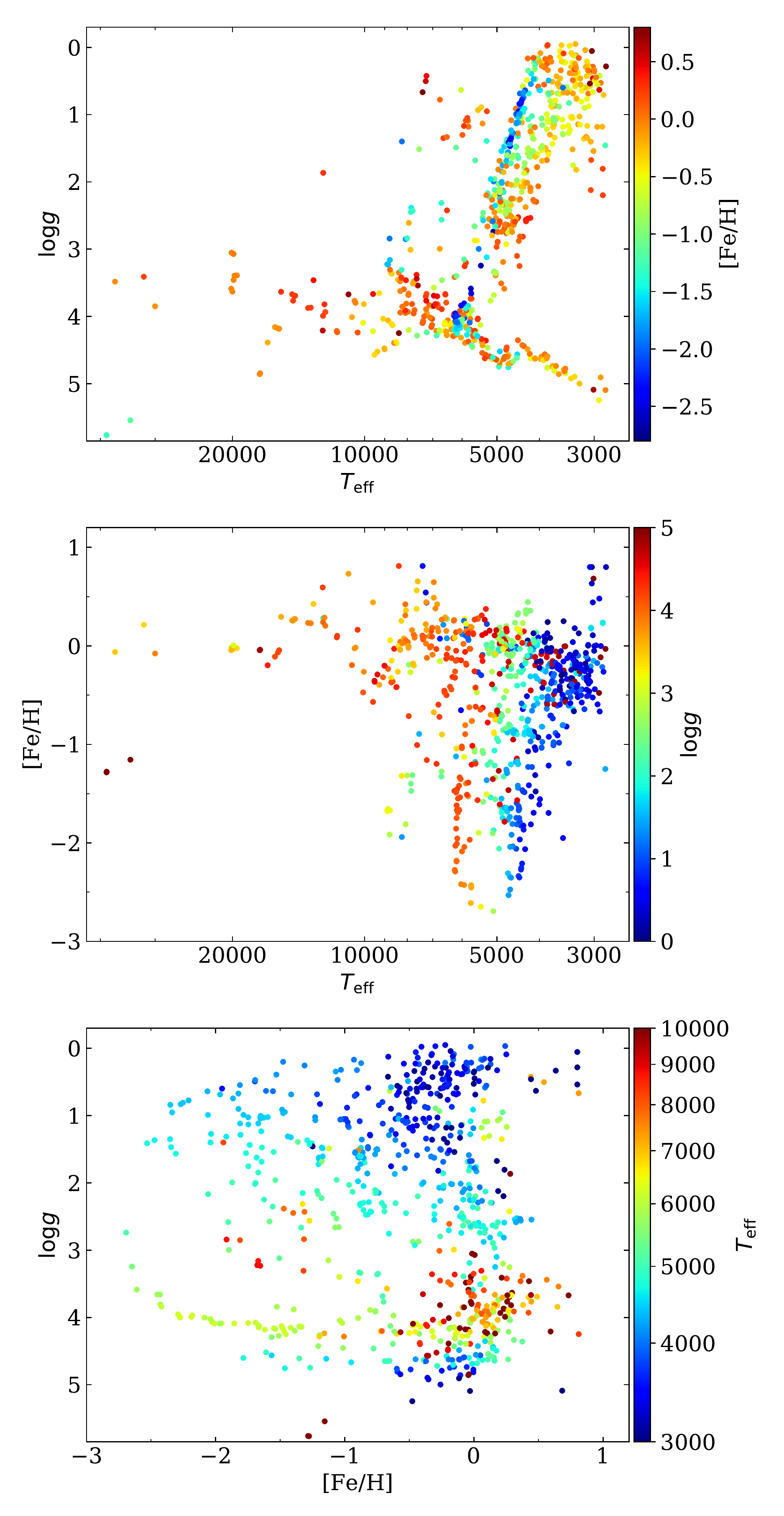}
\caption{Final parameters in this paper for XSL. Top panel: Hertzsprung-Russell diagram for XSL, colour-coded by [Fe/H]. Middle: Distribution in the \teff -- \feh plane, colour-coded by $\log g$. Bottom: Distribution in the \feh -- \logg plane, colour-coded by $\log \teff$. One star with a \logg error larger than 2.0 is not shown (carbon star [ABC89] Cir 18). }
    \label{fig:HR}
\end{figure}

Table~\ref{table:params} gives the derived atmospheric parameters for 754 observations of 616 stars. 
Their distribution in the \teff versus \logg, \teff versus \feh and \feh versus \logg planes are presented in Fig.~\ref{fig:HR}, to show that XSL contains a wide variety of stars. The library includes many normal dwarf and giant stars with a wide range in metallicity, and there are very hot stars, a significant number of AGB stars (with \teff $<$ 4000~K and \logg $<$ 1.0), Cepheids, RR Lyrae stars, post-AGB stars, and horizontal branch stars.

There is a twiddle around $\logg=4.8$ and $\teff=5000$ K, which is also seen in analyses of the Large Sky Area Multi-Object Fiber Spectroscopic Telescope (LAMOST) survey data by \citet{wuluo2011}, \citet{xiang2015}, and \citet{ho2017}, which use a similar analysis approach. This is certainly an artefact due to the limited number of stars with accurate gravity measurements in the ELODIE and MILES stellar libraries, which were used to create the interpolator. A fit of the solar spectrum returns a gravity that is slightly too low for the Sun. Improving the interpolator in this region is possible, but beyond the scope of the current work. The effect is worse at the transition where we accept the UVB solution alone instead of a combination of the UVB and VIS solutions (at 4500~K). The effect is however within the \logg uncertainties.

\subsection{External precision: Comparisons with the general literature} \label{sec:extcomp}

\begin{figure*}
\centering
\includegraphics[width=17cm]{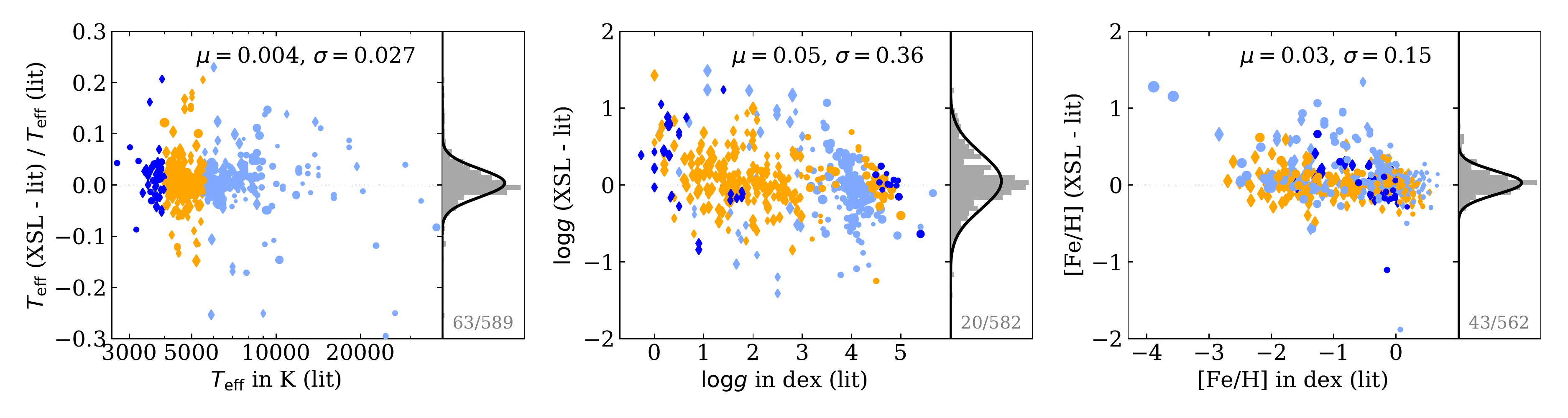}
\caption{Comparison between our adopted parameters and the literature compilation. The standard deviation and mean of the differences are written and the corresponding curve is shown. The number of clipped observations compared to the total number of literature values is presented in the lower right corner of each of the panels. The colour of the points relates to \teff; the size of the points to \feh, and giants and dwarfs (\logg limit 3.0) are indicated by diamonds and circles, respectively. In the \teff and \logg panels, two and four outliers, respectively, are not shown (various types of stars). }
    \label{fig:literature}
\end{figure*}

\begin{figure*}
\centering
\includegraphics[width=17cm,trim={1.0cm 0.5cm 1.0cm 0.0cm}]{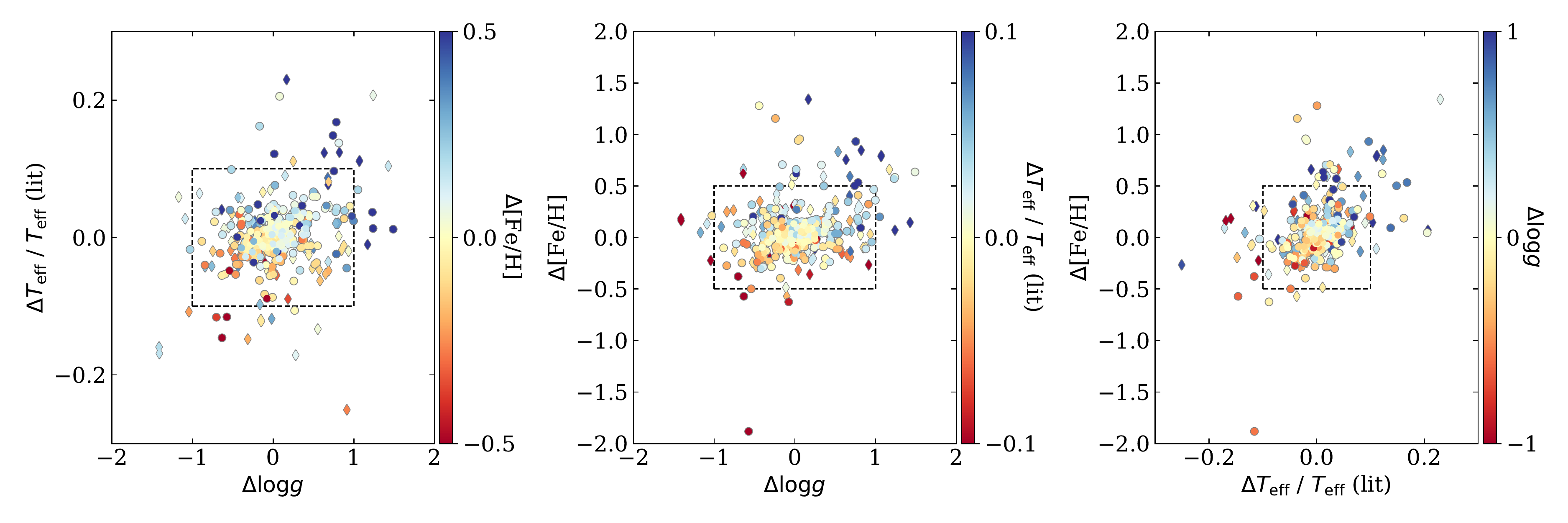}
\caption{Comparison of different $\Delta$s (our adopted parameter minus the literature value) for \teff, \logg, and \feh. Stars with multiple literature references are shown as circles, whereas stars with only one reference are shown as diamonds. The observations outside the boxes that have more than one literature reference are discussed in Sections~\ref{sec:litfeh}--\ref{sec:litlogg}.}
    \label{fig:literaturedeltas}
\end{figure*}

The most inclusive assessment of the precision of our determination is a comparison with measurements published in the literature. For this purpose we assembled a compilation based on the PASTEL catalogue \citep{soubiran16}, which compiles published atmosphere analysis from high resolution spectra and atmosphere models. We supplemented the catalogue by other series of measurements missing in PASTEL (but satisfying its selection criteria) and measurements that extend the boundaries of PASTEL.
These latter measurements include analyses of medium or low resolution spectra, using either synthetic spectra or observed stellar templates, and they also include photometric measurements. This compilation is a long-term effort, and to verify its completeness we checked the literature for each star using SIMBAD. The compilation, listed in Table~\ref{table:individuallit} (available in full only in electronic form in Vizier), contains 4123 measurements of the XSL stars from 445 references. Half of these measurements (2069) are in PASTEL.
For each star we computed a single set of parameters by averaging the available measurements, giving priority to those resulting from recent studies at high spectral resolution, and using other measurements only when no such measurements are available or when the high resolution measurements are not internally consistent. This procedure produces parameters classified in four quality classes. The Q0 class has the best values. They are based on high resolution spectroscopic original measurements published after 1985 and they are consistent within 6\% for \teff, 0.7 for \logg, and 0.4 for \feh. The Q1 class contains measurements from low resolution spectra and the older or less consistent high resolution spectra. The Q2 class also includes photometric measurements and Q3 are from compilations (from measurements not included otherwise).
The average literature parameters for the present sample, their standard deviation, the number of measurements used, and the quality parameters are presented in Table~\ref{table:litcompilation}.

In this compilation, \teff, \logg, and \feh are available, respectively, for 493, 487, and 466 of the 668 XSL stars. 
Including the repeated observations, 586 of the 786 measurements that we obtained can be compared to a measurement of \teff from the compilation. If we reject those of our measurements departing from the literature compilation by more than 10\% in \teff, or 1.0 and 0.5, respectively, in \logg or \feh, the 450 remaining measurements have a standard deviation of 3\% in \teff, and 0.33 and 0.16, respectively, in the two other parameters. Of these measurements 329 are for FGK stars, and for those the standard deviations are slightly lower. 

The difference between the literature and our measurements is presented in Fig.~\ref{fig:literature}. We computed standard deviations and means using the python \texttt{sigma\_clip} function with a clipping sigma of 3.0, where the number of stars rejected compared to the total available literature measurements is shown in the bottom right corner of the figure. 
The consistency between our measurements and the literature is better for the GK stars, and it degrades for the hotter, cooler, and lowest gravity stars.

In Fig.~\ref{fig:literaturedeltas}, we present our adopted parameter minus the literature value ($\Delta$) in combinations of all three parameters.
The dashed boxes represent deviations of 10\%, 1, and 0.5 on the three parameters, respectively, corresponding approximately to three times the clipped standard deviations. A mild correlation is seen along the expected temperature-metallicity degeneracy line (right panel), but this accounts only for a minor fraction of the variance.
The outliers in each projection (outside the dashed boxes) are most often discrepant on all the three parameters, 
and we note that their distribution is asymmetric. There are more outliers for which we measure \teff, \logg, and \feh values higher than in the literature than there are outliers with lower values. We do not think this asymmetry is physically significant.
In an attempt to identify cases or regions of the parameter space for which our method may fail,
we discuss below the outliers for which the literature reports more than one set of values considered reliable; these outliers are shown as circles outside the boxes.

\subsubsection{$\Delta \, \feh > 0.5$}\label{sec:litfeh}
There are 20 observations with a \feh difference larger than 0.5. Many of these observations belong to peculiar stars (that are not covered by the interpolator), for example two $\alpha$2~CVn variable stars (hot stars with abnormal chemical compositions), two extremely metal-poor stars (\feh $< -3.0$), and a very metal-poor cool subdwarf (\feh = $-2.5$), a brown dwarf, a Herbig~Ae/Be star with many emission lines, a hot (12\,000~K) metal-poor ($-0.4$) star, and a high proper motion cool flare star. 

There are also a few metal-poor stars (\feh $ < -1.5$) that ULySS has confused for a different evolutionary state, which also have a large \logg discrepancy. Two of these are horizontal branch stars that we place on the main sequence (HD074721, for which the \teff is also discrepant by 10\%) or on the giant branch (HD184266). Then there is a post-AGB star that is placed on the horizontal branch (HD116745). We carefully checked that our solutions were not trapped in a local minimum, and we suppose that the interpolator simply fails to reproduce these stars because they are not represented in the reference libraries.

Additionally, there are three observations (of two stars) that do not seem to have an obvious reason to be discrepant with the literature in \feh. The first star, HD184571, has a good fit in ULySS; the two arms give consistent solutions, but there are only low resolution spectroscopic \feh values available in the literature with a spread of 0.19. Therefore it is not clear whether our measurements or the literature measurements are better. 
The second star is CL*~NGC~330~ROB~A3 in the globular cluster NGC 330 in the Small Magellanic Cloud, which has two XSL observations. 
The \teff and \logg for this star are also discrepant with the literature, but the fits are very good and consistent with each other between the two observations. There may be a possibility that we are comparing this star to the wrong literature star in the cluster.
Finally, there are five stars for which the dispersion of \feh from different literature studies is $\geq$ 0.20 and a good comparison cannot be made. 

\subsubsection{$\Delta \, \teff > 10\%$}\label{sec:litteff}

There are 20 observations with a \teff difference larger than 10\%. Three stars are cool supergiants in the Magellanic Clouds with literature \teff~$< 3600$~K: [M2002]~LMC~150040, [M2002]~LMC~158646, and [M2002]~SMC~52334. Each of these stars has a high \teff dispersion in the literature (200--350 K), and each of the literature \teff values comes from comparing intermediate resolution spectroscopy to synthetic spectra; this kind of comparison is subject to many uncertainties \citep[e.g.][]{davies}. Another cool star, the red supergiant HD050877 with literature \teff = 3550 K, has multiple high resolution spectroscopic measurements, however their dispersion is 500~K. 

Then there is the warmer giant star HD058790 (literature \teff 5275~K) which is well fitted, consistent between the UVB and VIS, and there is no reason to believe its fitted parameters are wrong. The fitted \teff is 600~K cooler than the literature value, which is an average of two papers from 1989 and 1999 by the same group (\citealt{luck89}; \citealt{luck92}). The fitted \logg is also 0.7 lower than the literature. The spectrum of this star and its fit are very comparable to other XSL stars in this parameter region, therefore we assume our measurement is better. 

The spectrum of the F giant HD161770 is well fitted, but it disagrees with the literature \teff by 600~K. The literature \teff includes seven photometric measurements that are all relatively consistent with each other, and one spectroscopic measurement from \citet{bensby14} which is higher. Remarkably, our determined temperature agrees relatively well with that from that spectroscopic work. 

For HD188262, a warm giant, we have two observations, which in both cases show a large discrepancy between the UVB and VIS \teff values.  For both observations, we adopt the UVB solution which is 800~K warmer than the literature. However, the VIS solutions agree much better with the literature, although the literature values show a high dispersion of 600~K.  
 
There are two horizontal branch stars HD143459 and HD128801 with \teff $> 9000$~K, which both have a large difference in the \teff derived from the UVB or the VIS ($> 800$ K). The VIS barely contributes to our final adopted parameters in hot stars like these. In the case of HD128801, the UVB is closer to the literature \teff. For HD143459, the literature \teff is in between the UVB and VIS \teff. However, for both stars either the literature values have a high dispersion, or they are different measurements from one single paper (and thus not from independent methods).

There are also two observations of the hot subdwarf HD004539 that are off from the literature by 5000~K, where the literature \teff = 24570~K with a dispersion of 1200~K. However, all four measurements are from the same reference and therefore are also not independent.

Finally there are some discrepant stars which are peculiar stars. Examples are a post-AGB star which is an RV Tauri variable star, two Herbig Ae/Be stars, a Cepheid, an RR Lyrae star, and one hot variable supergiant. 

\subsubsection{$\Delta\,\logg > 1.0$}\label{sec:litlogg}

There are four observations with multiple literature measurements that agree with the literature for \teff and \feh within 10\% and 0.5, respectively, but are discrepant in \logg by more than 1.0. However, the literature dispersions in \logg for these stars are generally high. One is the very cool giant HD065354, which has a literature dispersion for \logg of 1.2. Secondly, there is another very cool giant [M2002] SMC 55188, for which we find a difference in \logg of 1.03. For this type of star there is very little reliable literature information available. Another  is the post-AGB star HD112374, however the literature dispersion is high with 0.8 in \logg. The fourth observation is one of the three observations for Feige 110, a hot subdwarf with \teff $= 39\,000$~K. The \logg values of the other two observations agree with the literature, but this one has a difference of 2, placing it at 3.5 instead of 5.5. It is however remarkable that the other two observations agree with the literature both in \logg and \teff because this type of star is not represented in the MILES library, and the success of our analysis is due to the theoretical extrapolation support spectra used to compute the interpolator. 

\begin{figure*}
\centering
\includegraphics[width=17cm]{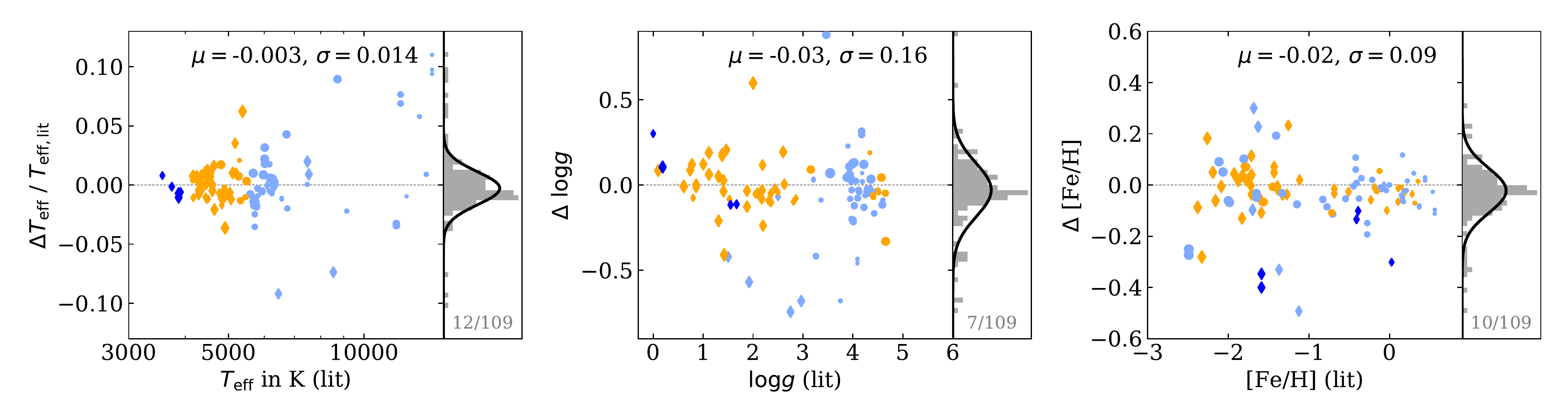}
\caption{Comparison between our adopted parameters and the parameters from \citet{wu}. One and three outliers are not shown for \logg and \feh, respectively, and one measurement at 29000 K is not shown (its $\Delta$ \teff / \teff$_\mathrm{lit}$ is $+0.027$.). The axes and symbols are explained in Fig.~\ref{fig:literature}.}
    \label{fig:literature_w11}
\end{figure*}

\begin{figure*}
\centering
\includegraphics[width=17cm]{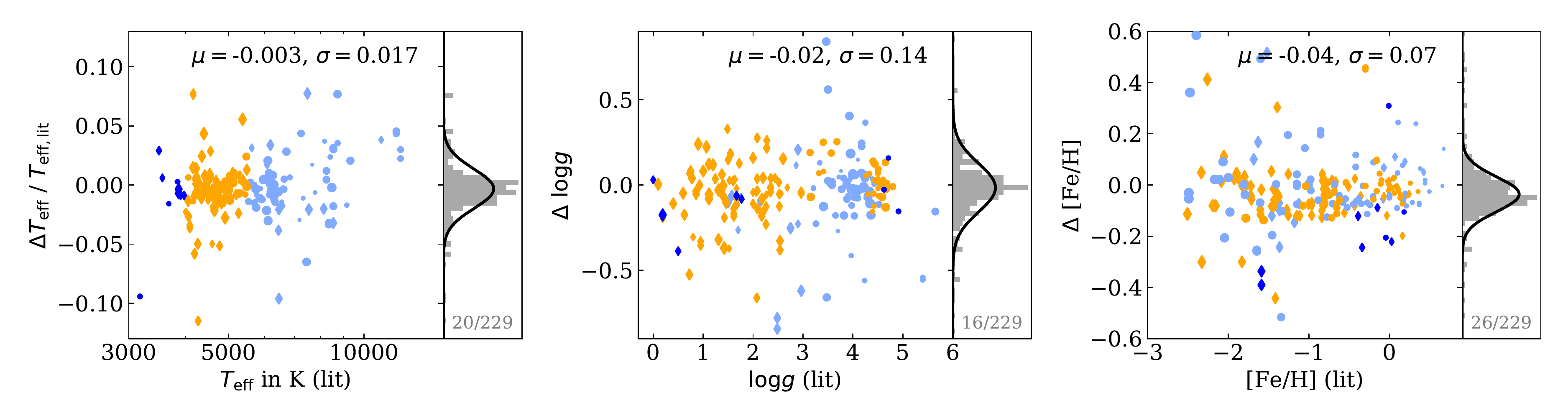}
\caption{Same as Fig.~\ref{fig:literature_w11}, but compared to \citet{prugniel11}. There are eight, four, and three outliers not shown for \teff, \logg, and \feh, respectively, and three additional measurements above 20000 K are not shown; they have differences between $-0.04$ and $-0.06$.}
    \label{fig:literature_p11}
\end{figure*}

\begin{figure*}
\centering
\includegraphics[width=17cm]{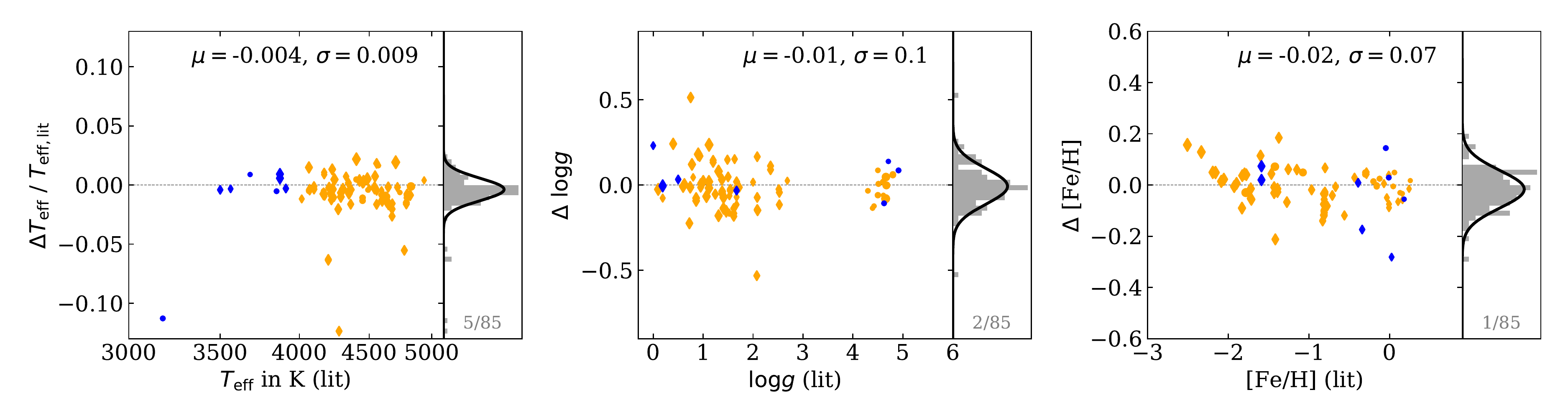}
\caption{Same as Fig.~\ref{fig:literature_w11}, but compared to \citet{sharma}. All measurements from XSL in common with that study are shown.}
    \label{fig:literature_s16}
\end{figure*}

\subsubsection{Summary of the literature comparison}

Overall, our measurements agree with the literature within 2.7\%, 0.36, and 0.15 for \teff, \logg, and \feh, respectively. The largest share of the dispersion is likely due to the inhomogeneity of the literature compilation. We generally trust our parameters for stars on the main sequence and those on the giant branch with $\teff > 3800$~K.
Compared to previous stellar libraries, in XSL there is a relatively large group of AGB stars and other giants cooler than 3800~K. For these stars, there is very little reliable information in the literature with which to compare.

We find that many of the parameters that are discrepant with the literature are for peculiar stars which are not well represented by the MILES interpolator, such as extremely metal-poor stars, brown dwarfs, stars with peculiar chemistry, and variable stars. Additionally, there are a significant number of stars for which the literature has a high dispersion itself. For some other stars, we have good fits and find no obvious reason for the discrepancy with the literature. There may be issues with the literature values themselves, or we might be comparing two different stars without knowing. Finally, we found that some of the stars discrepant with the literature also show discrepancies between our UVB and VIS measurements.

\subsection{Comparison with previous ULySS determinations}\label{sec:previousulyss}

To test the consistency of our approach, we compare our determinations with three previous studies using ULySS and the MILES or ELODIE interpolators. These comparisons also cannot be regarded as estimates of the external errors because these series may be affected by similar systematic because the same methods were used.

\subsubsection{ \citet{wu} }

 \citet{wu} used the ELODIE interpolator to determine parameters for the CFLIB (also known as the Indo-US library). Including the repeated observations, 109 measurements that we obtained can be compared to a measurement in that study; see Fig.~\ref{fig:literature_w11}. The standard deviations in \teff, \logg, and \feh are 1.4\%, 0.16, and 0.09, respectively (clipping with a sigma of 3.0). There do not appear to be any biases. \citet{wu} compared their determined parameters for FGK stars to many different literature compilations and determined robustly their external errors. They claim a precision of 43~K, 0.13, and 0.05 for \teff, \logg, and \feh, respectively. Our standard deviations with respect to \citet{wu} are slightly higher.
 
\subsubsection{\citet{prugniel11}}

\citet{prugniel11} used the first version of the MILES interpolator to determine uniform parameters for the MILES library. Including repeated observations, 229 observations can be compared to a measurement from that study; see Fig.~\ref{fig:literature_p11}. The standard deviations are 1.7\%, 0.14, and 0.07 in \teff, \logg, and \feh, respectively, although especially for \teff a large fraction of the stars are clipped from computing the standard deviation. There may be a small bias in \feh, where our measurements are on average 0.04 lower.

\subsubsection{\citet{sharma}}

Finally, we compare to the study by \citep[][hereafter S16]{sharma}, who improved the MILES interpolator for the cool stars (<~5000~K). This is the version of the interpolator we are using. For 85 observations in our sample we can compare to that study, which only contains cool stars; see Fig.~\ref{fig:literature_s16}. The standard deviations are 0.9\%, 0.1, and 0.07 in \teff, \logg , and \feh, respectively. There are only five observations with differences larger than 5\% in \teff; one of these is a flare star with a high proper motion (BD$+$19 5116B; our \teff, \logg and \feh 2900~K, 4.9, and $-0.12$ respectively), and the other three stars (one of these has two observations) are all giants between $4100-4500$~K in the same globular cluster NGC~6838: CL*~NGC~6838~AH~A9 (twice), NGC~6838~1009, and NGC~6838~1053. The S16 authors have fixed \feh to the cluster \feh while fitting \logg and \teff, whereas we fit \feh as a free parameter. This affects the final solutions, although the difference between our values and the adopted cluster \feh is small (0.08 -- 0.14). It is therefore unclear why the effective temperatures are discrepant. Additionally, S16 reanalysed their spectra without fixing \feh. The effective temperatures do not change much, so likely fixing \feh is not the only reason our measurements are discrepant with S16. CL*~NGC~6838~AH~A9 and NGC~6838~1053 are also in the literature compilation we compared to in Sec.~\ref{sec:extcomp} and have multiple \teff determinations from high resolution spectroscopy. Our \teff agrees better (within 50--100~K) with those literature values than the S16 \teff values, which are different from the literature by 200 and 400~K for CL*~NGC~6838~AH~A9 and NGC~6838~1053, respectively.


\section{Summary}\label{sec:summary}

In this paper we have presented uniformly derived stellar atmospheric parameters for XSL. We used two spectral regions from the UVB and VIS arm spectra of the instrument and performed full-spectrum fitting with the ULySS package and the MILES spectral interpolator.
We compared the solutions obtained for repeated observations of the same object to assess the accuracy of the internal errors. We compared the solutions obtained using the two spectral regions, and those obtained using another interpolator based on the ELODIE library to evaluate the systematics introduced by the method and the effects due to particular characteristics of individual stars.
This analysis indicates that for $\logg > 1$ and $4000 < \teff < 5500$ K the total errors are 0.9\% (or 44 K), 0.14,  and 0.06 for \teff, \logg, and \feh, respectively, and for lower temperatures the \feh total error increases to 0.1. For the hotter stars the estimated total errors are 2.6\%, 0.20, and 0.10 on the three parameters, and for $\logg < 1$, they are 2.1\%, 0.21, and 0.22. These errors are consistent with those previously reported in papers using the same method \citep{prugniel11,wu,koleva12}.

The XSL spectra in combination with these atmospheric parameters can be used to create spectral interpolators and stellar population models. This library has an important value because it has a large wavelength coverage obtained in a single observation, medium resolution (between R=8\,000 and 11\,000), and a wide coverage of the parameter space.


\begin{acknowledgements}
AA gratefully acknowledges funding by the Emmy Noether programme from the Deutsche Forschungsgemeinschaft (DFG). PhP and AL acknowledge financial support from the Programme National Cosmology and Galaxies (INSU/CNRS) in 2017 and 2018. AG is supported by the European Union FP7 programme through ERC grant number 320360. JFB and AV acknowledge support from grant AYA2016-77237-C3-1-P from the Spanish Ministry of Economy and Competitiveness (MINECO).
We thank Matthijs Dries for help with telluric corrections in XSL DR2. 
We thank Paula Coelho for her useful comments on a draft of this work. 
This research has made use of the SIMBAD database, operated at CDS, Strasbourg, France.
\end{acknowledgements}


\bibliographystyle{aa}
\bibliography{XSLparam.bib}   


\clearpage

\appendix
\onecolumn
\section{Tables with derived and literature stellar parameters}
\longtab{}{
\setlength{\tabcolsep}{2.0pt}
\begin{longtable}{lrrrrrrrrrrrrrr}
\caption{\label{table:params}Derived atmospheric parameters of the XSL stars}\\
\hline\hline
Name &
Obs\tablefootmark{a} &
\multicolumn{3}{c}{UVB solution}&
\multicolumn{3}{c}{VIS solution}&
\multicolumn{6}{c}{Adopted parameters} &
flag\tablefootmark{b} \\
 &
 &
$T_{\rm eff}$ & $\log g$ & [Fe/H] & 
$T_{\rm eff}$ & $\log g$ & [Fe/H] & 
$T_{\rm eff}$ & err & $\log g$ & err & [Fe/H] & err  & \\
&
&
(K)&($[cm\,s^{-2}]$)&([Sun])&
(K)&($[cm\,s^{-2}]$)&([Sun])&
\multicolumn{2}{c}{(K)}            & \multicolumn{2}{c}{($[cm\,s^{-2}]$)} & \multicolumn{2}{c}{([Sun])} & \\
\hline

\endfirsthead
\caption{continued.}\\
\hline\hline
Name &
Obs\tablefootmark{a} &
\multicolumn{3}{c}{UVB solution}&
\multicolumn{3}{c}{VIS solution}&
\multicolumn{6}{c}{Adopted parameters} &
flag\tablefootmark{b}\\
 &
 &
$T_{\rm eff}$ & $\log g$ & [Fe/H] & 
$T_{\rm eff}$ & $\log g$ & [Fe/H] & 
$T_{\rm eff}$ & err & $\log g$ & err & [Fe/H] & err  & \\
&
&
(K)&($[cm\,s^{-2}]$)&([Sun])&
(K)&($[cm\,s^{-2}]$)&([Sun])&
\multicolumn{2}{c}{(K)}            & \multicolumn{2}{c}{($[cm\,s^{-2}]$)} & \multicolumn{2}{c}{([Sun])} & \\
\hline
\endhead
\hline
\endfoot

2MASS J17535707$-$2931427 & 247 & 4723 & 2.03 & $-$0.04 & 4722 & 2.31 & $-$0.02 & 4723 & 37 & 2.03 & 0.15 & $-$0.04 & 0.05 &  \\ 
2MASS J18024572$-$3001120 & 255 & 4495 & 1.57 & 1.00 & 3296 & 1.82 & $-$0.08 & 3296 & 35 & 1.82 & 0.82 & $-$0.28 & 2.27 &  \\ 
2MASS J18024611$-$3004509 & 256 & 2700 & 5.71 & 1.00 & 3096 & 0.79 & 0.21 & 3096 & 67 & 0.69 & 0.30 & 0.01 & 0.25 &  \\ 
2MASS J18025277$-$2954335 & 257 & 3328 & 1.44 & 0.39 & 3368 & 0.33 & $-$0.25 & 3368 & 56 & 0.23 & 0.20 & $-$0.45 & 0.20 &  \\ 
2MASS J18032525$-$2959483 & 169 &   &   &   & 3270 & 0.50 & $-$0.32 & 3270 & 56 & 0.40 & 0.22 & $-$0.52 & 0.24 &  \\ 
2MASS J18033716$-$2954227 & 152 & 4237 & 2.49 & 0.34 & 4297 & 2.56 & 0.52 & 4250 & 30 & 2.55 & 0.15 & 0.45 & 0.05 &  \\ 
2MASS J18040638$-$3010497 & 153 & 3354 & 1.19 & 0.57 & 3384 & 0.39 & $-$0.04 & 3384 & 56 & 0.29 & 0.20 & $-$0.24 & 0.19 &  \\ 
2MASS J18042244$-$3000534 & 154 & 2958 & 0.88 & $-$0.09 & 3240 & 1.03 & $-$0.08 & 3240 & 41 & 1.03 & 0.21 & $-$0.28 & 0.20 &  \\ 
... & ... & ... & ... & ... & ... & ... & ... & ... & ... & ... & ... & ... & ... & \\

\hline
\end{longtable}
\tablefoot{
\tablefoottext{a}{XSL observation identifier. Internal unique number for an observation; repeated observations of the same target have different numbers, so this number should not be used to identify a star.}
\tablefoottext{b}{Flag for stars that show large differences with their literature stellar parameters (if there are at least two literature measurements available), as discussed in Sect.~\ref{sec:extcomp}. Carbon stars studied by \citet{gonneau16,gonneau17} have also been flagged. The flags should be combined with the quality flags for the spectra from the XSL DR2 paper (Gonneau et al. in prep).} 
This table is available in its entirety at the CDS.
} 
} 

\longtab{}{
\setlength{\tabcolsep}{4pt}
\begin{longtable}{lrrrrrrrrrrr}
\caption{\label{table:individuallit} 4123 individual atmospheric parameters from the literature compilation$^{a}$}\\
\hline\hline
Name & J2000 & $T_{\rm eff}$ & err & $\log g$ & err & [Fe/H] & err & r & c & m & bibcode \\ 
&  & \multicolumn{2}{c}{(K)} & \multicolumn{2}{c}{($[cm\,s^{-2}]$)} &  \multicolumn{2}{c}{([Sun])}  & & & \\
\hline

\endfirsthead
\caption{continued.}\\
\hline\hline
Name & J2000 & $T_{\rm eff}$ & err & $\log g$ & err & [Fe/H] & err & r & c & m & bibcode \\ 
& & \multicolumn{2}{c}{(K)} & \multicolumn{2}{c}{($[cm\,s^{-2}]$)} &  \multicolumn{2}{c}{([Sun])}  & & & \\
 
\hline
\endhead
\hline
\endfoot

2MASS J18033716$-$2954227 &  J180337.1$-$295422  & 4300   &   &   1.67  &   &    0.54 & 0.37 & L & O & A & 2008A\&A...486..177Z \\
2MASS J18351799$-$3428093 &  J183518.0$-$342809  & 4750   &   &   1.98  &   &   $-$0.19 & 0.12 & L & O & A & 2008A\&A...486..177Z \\
2MASS J18352206$-$3429112 &  J183522.1$-$342911  & 4850   &   &   2.03  &   &    0.01 & 0.14 & L & O & A & 2008A\&A...486..177Z \\
2MASS J18352834$-$3444085 &  J183528.3$-$344409  & 4900   &   &   2.00  &   &   $-$1.40 & 0.18 & L & O & A & 2008A\&A...486..177Z \\
2MASS J18355679$-$3434481 &  J183556.8$-$343448  & 4350   &   &   1.81  &   &   $-$0.67 & 0.10 & L & O & A & 2008A\&A...486..177Z \\
BD+01 2916              &  J142145.3+004659  & 4150   &   &   0.1   &   &   $-$1.99 &      & H & O & A & 2000ApJ...544..302B \\
BD+01 2916              &  J142145.3+004659  & 4150   &   &   0.1   &   &   $-$1.99 &      & H & O & A & 1996AJ....111.1689P \\
BD+01 2916 & J142145.3+004659 &  &  &  &  & $-$1.48 &  &   H &  O &  A &  1983ApJ...271L..75L \\
... & ... & ... & ... & ... & ... & ... & ... & ... & ... & ... &  ... \\

\hline
\end{longtable}
\tablefoot{
\tablefoottext{a}{These measurements are compiled as described in Sec.~\ref{sec:extcomp}. There are three quality measures, indicated in columns r, c and m. The column r indicates the resolution class, high (H), low (L) or photometry (P). The column c describes the origin, original measurement (O) or compilation (C). Column m describes the method, referenced to model atmosphere (A), isochrones (I) or empirical calibration (T). This table is available in its entirety at the CDS.} \\
} 
} 

\longtab{}{
\setlength{\tabcolsep}{5pt}
\begin{longtable}{lrrrrrrrrrrrr}
\caption{\label{table:litcompilation} Mean atmospheric parameters from the literature comparison for 487 stars$^{a}$}\\
\hline\hline
Name & \multicolumn{4}{c}{$T_{\rm eff}$ (K)} & \multicolumn{4}{c}{$\log g$ ($[cm\,s^{-2}]$)} & \multicolumn{4}{c}{[Fe/H] ([Sun])} \\ 
 & $\mu$ & $\sigma$ & $n$ & $Q$ & $\mu$ & $\sigma$  & $n$ & $Q$ & $\mu$ & $\sigma$  & $n$ & $Q$ \\
\hline

\endfirsthead
\caption{continued.}\\
\hline\hline
Name & \multicolumn{4}{c}{$T_{\rm eff}$ (K)} & \multicolumn{4}{c}{$\log g$ ($[cm\,s^{-2}]$)} & \multicolumn{4}{c}{[Fe/H] ([Sun])} \\ 
 & $\mu$ & $\sigma$ & $n$ & $Q$ & $\mu$ & $\sigma$  & $n$ & $Q$ & $\mu$ & $\sigma$  & $n$ & $Q$ \\
\hline
\endhead
\hline
\endfoot

2MASS J18033716$-$2954227 & 4300 &  & 1 & 1 & 1.67 &  & 1 & 1 & 0.54 &  & 1 & 1 \\
2MASS J18351799$-$3428093 & 4750 &  & 1 & 1 & 1.98 &  & 1 & 1 & $-$0.19 &  & 1 & 1 \\
2MASS J18352206$-$3429112 & 4850 &  & 1 & 1 & 2.03 &  & 1 & 1 & 0.01 &  & 1 & 1 \\
2MASS J18352834$-$3444085 & 4900 &  & 1 & 1 & 2.00 &  & 1 & 1 & $-$1.40 &  & 1 & 1 \\
2MASS J18355679$-$3434481 & 4350 &  & 1 & 1 & 1.81 &  & 1 & 1 & $-$0.67 &  & 1 & 1 \\
BD+01 2916 & 4167 & 29 & 3 & 0 & 0.10 & 0.00 & 3 & 0 & $-$1.93 & 0.11 & 3 & 0 \\
BD+03 2688 & 4300 &  & 1 & 0 & 0.00 &  & 1 & 0 & $-$1.42 &  & 1 & 0 \\
BD+04 2466 & 5031 & 137 & 8 & 0 & 1.61 & 0.54 & 3 & 0 & $-$1.99 & 0.12 & 4 & 0 \\
... & ... & ... & ... & ... & ... & ... & ... & ... & ... & ... & ... & ... \\

\hline
\end{longtable}
\tablefoot{
\tablefoottext{a}{These measurements are compiled as described in Sec.~\ref{sec:extcomp}. Column names $\mu$, $\sigma$, $n$ and $Q$ refer to the average of the measurements in the literature, their standard deviation, the number of included measurements and the quality of the combined measurement (as described in the main text), respectively. }
This table is available in its entirety at the CDS. \\
} 
} 

\end{document}